\documentclass[pre,aps,floats,superscriptaddress,floatfix,twocolumn]{revtex4-2}
\usepackage{amssymb,amsmath}
\usepackage{mathtools}
\usepackage{graphicx}
\usepackage{psfrag}
\usepackage{color}
\usepackage{soul}
\usepackage{dcolumn}
\usepackage[utf8]{inputenc}
\usepackage{hyperref}

\usepackage{bm}
\usepackage[normalem]{ulem}
 \hypersetup{
    colorlinks=true,
    citecolor=blue,
    }
 \usepackage{amsmath}
 \usepackage{tabularx}
 \usepackage{array}
 \newcolumntype{Y}{>{\centering\arraybackslash}X}
\renewcommand{\arraystretch}{1.5}  % Increase row height for better spacing
 
\begin{document}

 \title{Stationary densities and delocalized domain walls in asymmetric exclusion
processes competing for finite pools of resources}
 \author{Sourav Pal}\email{isourav81@gmail.com}
\affiliation{Theory Division, Saha Institute of
Nuclear Physics, a CI of Homi Bhabha National Institute, 1/AF Bidhannagar, Calcutta 700064, West Bengal, India}
\author{Parna Roy}\email{parna.roy14@gmail.com}
\affiliation{Shahid Matangini Hazra Government General Degree College for Women, Purba Medinipore 721649, West Bengal, India}
\author{Abhik Basu}\email{abhik.123@gmail.com, abhik.basu@saha.ac.in}
\affiliation{Theory Division, Saha Institute of
Nuclear Physics, a CI of Homi Bhabha National Institute, 1/AF Bidhannagar, Calcutta 700064, West Bengal, India}

\begin{abstract}
 We explore the stationary densities and domain walls in the steady states of a pair of asymmetric exclusion processes (TASEP) antiparallelly coupled to two particle reservoirs without any spatial extent by using the model in Haldar et al., Phys. Rev. E {\bf 111}, 014154 (2025). We show that  the model admits a pair of {\em delocalized} domain walls, which exist for some choices of the model parameters that define the effective entry and exit rates into the TASEP lanes. Surprisingly, in the parameter space spanned by these model parameters, the region corresponding to delocalized domain walls covers an {\em extended} region, in contrast to the delocalized domain walls that appear only along a line in  the relevant parameter space of the other known variants of TASEP. This implies large fluctuations in the TASEP particle numbers even in the thermodynamic limit that can be found over a range of the control parameters. The corresponding phase diagrams in the plane of the control parameters have different topology from those for an open TASEP or other models with multiple TASEPs connected to two reservoirs.
\end{abstract}

\maketitle

 \section{INTRODUCTION}
 \label{introduction}

 Driven diffusive systems~\cite{driv1,driv2,driv3,driv4} can show highly nontrivial
properties even with purely local dynamics in even one dimension (1D). A paradigmatic
example is the totally asymmetric simple exclusion process
(TASEP)~\cite{krug-prl,derrida1,derrida2,derrida3}. It consists of a 1D lattice
with open boundaries. A TASEP's dynamics are stochastic and subject to exclusion at all sites, meaning that it can only hold a single particle per site at a time. A particle enters at one of the boundaries at a given rate, hops unidirectionally to the subsequent sites at a rate of unity until it reaches the other end, and then leaves the system at another specified rate. The TASEP shows three different phases as a function of the entry and exit rates at the boundary: the low-density phase is given by steady-state densities in bulk that are less than half, the high-density phase  by densities greater than half, and the maximal current phase with densities of just half.

The TASEP was first put forth as a minimal model to describe proteins synthesis in biological cells~\cite{macdonald}. Subsequently it was found to display boundary-induced phase transitions~\cite{krug-prl,derrida1,derrida2}. TASEP has been generalized to model various phenomena in biological and social systems, which has resulted in the identification of new phases and novel collective phenomena. For example, it has been found that traffic jams can be caused by the particle exchange between the environment and the TASEP lane~\cite{ef-lktasep-prl,ef-lktasep-pre}, which has been confirmed experimentally~\cite{traffic-exp}. Further investigation of these findings revealed that molecular motors belonging to the kinesin-8 family can govern the kinetics of microtubule depolymerization, which in turn leads to microtubule length regulation~\cite{melbinger1,melbinger2}. Furthermore, systems with two TASEP lanes coupled by infrequent particle-switching events have been found to exhibit complex phenomena, e.g., domain wall delocalization and traffic jams~\cite{reichenbach1,reichenbach2,reichenbach3}.

TASEP has also been investigated on simple networks, examining the requirements for the existence of the different phases and domain walls~\cite{net1,net2,net3} on different segments of the networks. These studies have been inspired by the dynamics of molecular motors traversing intricate networks of microtubules in real cells~\cite{net-exp}. In yet another generalization of TASEP, the interaction of particle number conservation with TASEP on an inhomogeneous ring has been explored and found to produce macroscopically nonuniform stable states of various kinds~\cite{def1,def2,sarkar-basu,banerjee-sarkar-basu,def5,def6,def7,def8,def9,def10}. Combining TASEP with the symmetric exclusion process (SEP) provides yet another intriguing generalization of TASEP. Both half-closed~\cite{half1,half2} and fully open systems~\cite{half3} have been used in this context. These studies have produced some noteworthy findings, including complex space-dependent stationary states in both TASEP and SEP lanes~\cite{half3}, self-organized temporal patterns~\cite{half2}, and tip localization of motors~\cite{half1}. In approaches complementary to Refs.~\cite{ef-lktasep-prl,ef-lktasep-pre},  a TASEP lane is considered to be connected to an embedded three-dimensional (3D) particle reservoir via particle exchanges in order to mimic the coupled system of molecular motors and microtubules. For instance, Refs.~\cite{klumpp1,klumpp2,ciandrini,dauloudet} have examined the effects of diffusion of the motors on the filament's steady states (TASEP), providing physical insights into the nontrivial steady states that result from combining 1D driven (TASEP) and 3D equilibrium (diffusion) processes. In addition to these investigations, but adhering strictly to a 1D description, Ref.~\cite{klumpp3}  examined a 1D open model with serially coupled diffusive and driven segments and the alterations that resulted in the pure open TASEP phase diagram. TASEP has also been used as a minimal model for traffic flow along network of roads~\cite{Chowdhury-Santen-Schadschneider,Schadschneider}.

In this work, we study a distinct class of TASEP models that are very different from those discussed above. These models place a limit on the total number of particles in the system of the TASEP connected to a particle reservoir at both ends rather than having open boundaries. These models are inspired by the fact that 
in the real world, limited amount of resources are available for use. For instance, finite availability of ribosomes in mRNA translocation for protein synthesis in eukaryotic cells~\cite{reser1}, or a finite number of vehicles in a closed network~\cite{ha-den}. This necessitates the extension of an unconstrained TASEP into closed systems that are composed of TASEP(s) connected to particle reservoir(s) with overall particle conserving dynamics. These models can also describe traffic dynamics, where a network of roads having traffic flow contains a limited fixed number of vehicles~\cite{traff1};  see Ref.~\cite{traff2} for similar studies. In these models, the primary effect of the finite availability of resources is that the effective entry rate of the particles to the TASEP, e.g., the actual protein synthesis taking place, sensitively depends upon the available resources. As the resource supply go up (reduce), protein synthesis rates increase (decrease). The main qualitative question in each of these studies is how the presence of a limited pool of accessible particles alters the steady-state densities and currents compared to those of open TASEPs with an unconstrained particle supply. From the standpoint of nonequilibrium statistical mechanics, these models serve as minimal models for ``boundary-induced nonequilibrium phase transitions with a conservation law''. In the recent past, there have been several studies focusing on aspects of the stationary states in TASEPs with finite resources systems and are composed of single TASEP or more~\cite{ha-den,reser1,ciandrini,reser2,reser3,cook-dong-lafleur} connected to a particle reservoir. How the phase diagrams and stationary densities and domain walls are controlled by the finite resources is the major theme of these works. In all these previous studies, the effective entry rate depends on the instantaneous reservoir population, however the particle exit rates from the TASEP lane to the reservoir are considered a constant, independent of reservoir population. Moreover, only one reservoir is employed in all these studies. Considering reservoir population-dependent exit rates allow one to study the ``reservoir crowding effect'', proposed and studied in Ref.~\cite{astik-parna}. The role and effects of distributed resources, as opposed to a single point reservoir, has also been explored in the recent past. These studies include research on how two reservoirs interact with two TASEPs~\cite{arvind2,sourav-1,sourav-2,astik-erwin} and how reservoirs interact in a bidirectional system~\cite{arvind-bi}. See Ref.~\cite{seppa} for a related study on a TASEP with $K$-exclusion, which generalizes the conventional TASEP and allows for up to $K$ particles at each site.%, and Ref.~\cite{sourav-4} for a study on the appearance of universal and nonuniversal domain walls in a model for TASEP with finite resources and a point defect.

In this article, we explore the stationary densities and the domain walls in a model with two asymmetric exclusion processes (TASEP) coupled antiparallelly to two particle reservoirs. This model was originally considered in Ref.~\cite{astik-erwin}. By considering a reduced parameter space distinct from that considered in Ref.~\cite{astik-erwin}, we show that the stationary densities in the two TASEP lanes are statistically identical, which can be spatially uniform or nonuniform. In the latter case, we find a pair of {\em delocalized domain walls} (DDW), one in each of the TASEP lanes, spanning completely or partially the lanes. The phase diagrams of the system in this reduced parameter space as functions of the control parameters are obtained, which are quite different from those found in Ref.~\cite{astik-erwin}, or for other models of TASEPs with finite resources. The most striking feature is that the phase space region in the plane of the control parameters where the DDWs are found is {\em extended}, as opposed to just a line found usually in other models of TASEP with or without open boundaries, or even with finite resources; see, e.g., ~\cite{sarkar-basu,astik-parna,astik-erwin,blythe} among others. Apart from its theoretical novelty, this has profound phenomenological implications: it means that for the control parameters in an extended region of the phase space, the stationary densities actually have large fluctuations, since DDWs are nothing but {\em moving} density shocks. Surprisingly, there is never population imbalances between the two reservoirs, unlike in Ref.~\cite{astik-erwin}. Furthermore, even when the TASEP lanes display large fluctuations  in the thermodynamic limit, the relative fluctuations in the {reservoir populations} decreases monotonically with larger TASEP size, suggesting vanishing relative reservoir number fluctuations in the thermodynamic limit. The remainder of this article is organized as follows: In Section~\ref{model}, we discuss the model. Then in Section~\ref{mft}, we outline the construction of the MFT for this model. Next in Section~\ref{pd}, we present the phase diagrams in the plane of the control parameters. Then in Section~\ref{steady-state-density}, we use MFT to obtain the stationary densities in the different phases and corroborate them by extensive MCS studies. Next, we discuss the absence of nonidentical stationary densities in the two TASEP lanes in Sec.~\ref{exclusion-of-asym-phases}. Then in Sec.~\ref{pb}, we calculate the different phase boundaries,  and the nature of the phase transitions in the model. We conclude by providing a short summary of our results and future prospects in Section~\ref{conclusion}. Additional results and some technical details are available in Appendix~\ref{ph-and-pd} and Appendix~\ref{ld-hd-mc-den} for interested readers.

 \section{THE MODEL}
 \label{model}

\begin{figure}[!h]
 \centerline{
 \includegraphics[width=\linewidth]{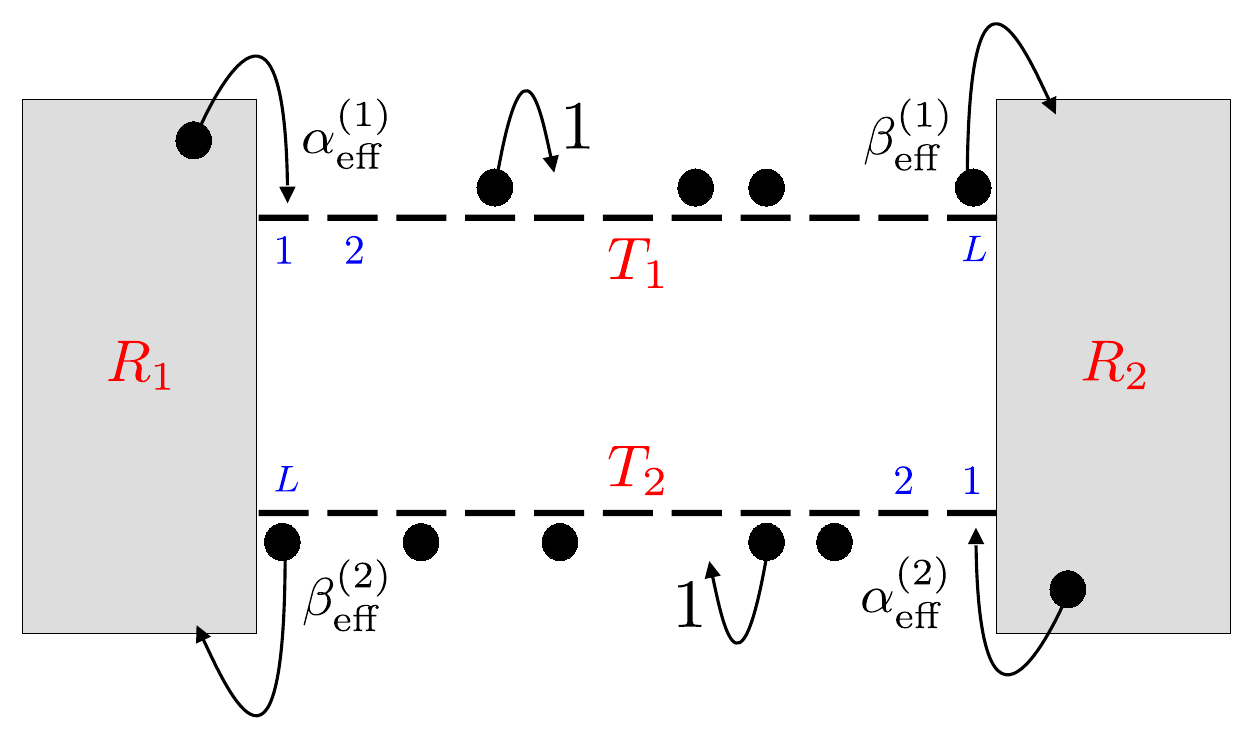}}
 \caption{\textbf{Schematic model diagram}: Two lattices $T_{1}$ and $T_{2}$ with $L$ sites in each are connected to two particle reservoirs $R_{1}$ and $R_{2}$ containing finite number of particles $N_{1}$ and $N_{2}$, respectively. Particles move following the TASEP update rules (1)-(5), see text. Unlike the open TASEP, entry and exit rates are dynamically controlled by instantaneous reservoir populations, see Eqs.~(\ref{eff-rates-T1})-(\ref{f-and-g}).}
 \label{model-diagram}
 \end{figure}

 We now revisit the construction and dynamics of the model proposed and studied in Ref.~\cite{astik-erwin} that we use for our work. The model is composed of two 1D lattices, $T_{1}$ and $T_{2}$, of equal length exhibiting dynamical rules of TASEP which are connected to two reservoirs, $R_{1}$ and $R_{2}$,  with finite particle-carrying capacity at their ends, see Fig.~\ref{model-diagram} for the schematic diagram. The TASEP lanes are of equal length $L$ with their sites labeled as $j=1,2,...,L$. Assumed to be point reservoirs without any spatial extent or internal dynamics, $R_{1}$ and $R_{2}$ respectively have $N_{1}$ and $N_{2}$ number of particles in them that are functions of time. The particles in the system interact with each other only through hard-core exclusion, meaning that each site on both TASEP lanes can be occupied by no more than one particle at a given time. Thus, each TASEP site can be in either of two states: empty or occupied. We follow the dynamics prescribed in Ref.~\cite{astik-erwin}. We briefly describe that below for the convenience of the reader.
 
\begin{itemize}
\item[(1)] At each time step, one site is randomly selected from each of $T_{1}$ and $T_{2}$.
\item[(2)] If the selected site is $j=1$ and empty, a particle enters from $R_{1}$($R_{2}$) into $T_{1}$($T_{2}$) with rate $\alpha_\text{eff}^{(1)}$($\alpha_\text{eff}^{(2)}$).
\item[(3)] If the site is $j=L$ and occupied, a particle exits from $T_{1}$($T_{2}$) to $R_{2}$ ($R_{1}$) with rate $\beta_\text{eff}^{(1)}$($\beta_\text{eff}^{(2)}$).
\item[(4)] The effective rates depend on the instantaneous reservoir populations $N_1$ and $N_2$ [see Eqs.~(\ref{eff-rates-T1})-(\ref{f-and-g}) below].
\item[(5)] If $1 < j < L$, a particle hops to the next site $j+1$ -- rightward in $T_{1}$, leftward in $T_{2}$ -- at rate 1, if site $j+1$ is empty; see Fig.~\ref{model-diagram}.
\end{itemize}

 The total particle number in the system, including the reservoirs and the TASEP lanes, is denoted by $N_{0}$, which is strictly \textit{conserved}. Particle number conservation (PNC) reads
\begin{equation}
 \label{pnc equation}
  N_{0}=N_{1}+N_{2}+\sum_{i=1}^{2}\sum_{j=1}^{L}n_{j}^{(i)},
 \end{equation}
where $n_{j}^{(i)}$ is the occupation number of the $j$-th site of $i$-th lane with $n_{j}^{(i)}=0(1)$ for empty(occupied) sites.

 As in Ref.~\cite{astik-erwin}, the effective entry and exit rates of particles into and out of the TASEP lanes are parametrized by the parameters $\alpha_{i}$ and $\beta_{i}$ ($i=1,2$), which are positive definite, as follows:
\begin{align}
\alpha_\text{eff}^{(1)} &= \alpha_1 f(N_1), \quad & \beta_\text{eff}^{(1)} &= \beta_1 g(N_2), \label{eff-rates-T1} \\
\alpha_\text{eff}^{(2)} &= \alpha_2 f(N_2), \quad & \beta_\text{eff}^{(2)} &= \beta_2 g(N_1). \label{eff-rates-T2}
\end{align}
The functions $f(N_{i})$ and $g(N_{i})$ must be monotonically increasing and decreasing in $N_{i}$, respectively, since an increasing reservoir population promotes particle entry into the associated TASEP lane but inhibits their exit. Various forms of such rate functions satisfying monotonicity have been proposed in the literature~\cite{greulich-ciandrini,reser1,cook-dong-lafleur,astik-parna,arvind1,arvind2,sourav-1,sourav-2}. For reasons of simplicity, we use the forms in Ref.~\cite{astik-erwin}
\begin{equation}
  \label{f-and-g}
  f(N_{i}) = \frac{N_{i}}{L}, \quad g(N_{i}) = 1 - \frac{N_{i}}{L}.
\end{equation}

 We define the filling factor $\mu$ as
\begin{equation}
 \label{filling-factor}
  \mu=\frac{N_{0}}{2L}
 \end{equation}
 which serves as a measure of overall resource availability. In this model, $\mu$ lies within the range $0 \le \mu \le 2$, corresponding to an empty ($\mu = 0$), fully occupied ($\mu = 2$), and half-filled ($\mu = 1$) system.

To reduce the number of control parameters, we consider the symmetric case $\alpha_{1} = \alpha_{2} \equiv \alpha$ and $\beta_{1} = \beta_{2} \equiv \beta$. These choices are distinct from those in Ref.~\cite{astik-erwin}, and allow us to explore a reduced parameter space different  from the one considered in Ref.~\cite{astik-erwin}. The model is thus governed by three parameters:  $\alpha$, $\beta$, and $\mu$.

 \section{MEAN-FIELD THEORY}
 \label{mft}

  In this section, we analyze the model at the mean-field~\cite{blythe} level, following the logic developed in Ref.~\cite{astik-erwin}.

Neglecting correlations in the mean-field theory (MFT), the time-averaged density $\rho_{j}^{(i)}\equiv\langle n_{j}^{(i)}  \rangle$ of bulk sites \(1 < j < L\) of $i$-th lane evolve according to the following equations of motion:
\begin{equation}
\label{bulk-equation}
\partial_t \rho_j^{(i)} = \rho_{j-1}^{(i)} (1 - \rho_j^{(i)}) - \rho_j^{(i)} (1 - \rho_{j+1}^{(i)}),
\end{equation}
with boundary conditions:
\begin{eqnarray}
\rho_1^{(i)} &=& \alpha_\text{eff}^{(i)}, \label{bc-entry} \\
\rho_L^{(i)} &=& 1 - \beta_\text{eff}^{(i)}. \label{bc-exit}
\end{eqnarray}
Equations~(\ref{bulk-equation})-(\ref{bc-exit}) are invariant under the following set of transformations (see Appendix~\ref{ph-and-pd} for details):
\begin{eqnarray}
  \alpha &\leftrightarrow& \beta, \label{tr-2} \\
  \rho_{j}^{(i)} &\leftrightarrow& 1 - \rho_{L - j + 1}^{(i)}, \label{tr-1}\\
  \mu &\leftrightarrow& 2-\mu. \label{tr-3}
\end{eqnarray}
In order to set up a continuum description, we introduce a  spatial variable \(x = j \epsilon\), with lattice spacing \(\epsilon = 1/L\). In the limit \(L \gg 1\), \(x\) becomes quasi-continuous, and the density \(\rho^{(i)}(x,t)\) is labeled accordingly.

In the steady state, the mean-field particle current
\[
J^{(i)}[\rho^{(i)}(x)] = \rho^{(i)}(x)(1 - \rho^{(i)}(x))
\]
 becomes constant. Denoting the constant currents in lanes \(T_1\) and \(T_2\) as \(J^{(1)}\) and \(J^{(2)}\) respectively, the flux balance equation relating these currents to the reservoir occupation numbers \(N_1\) and \(N_2\) reads:
\begin{equation}
\label{flux-balance-equation}
\frac{dN_1}{dt} = J^{(2)} - J^{(1)} = -\frac{dN_2}{dt}.
\end{equation}
In the steady state, reservoir populations are time-independent and Eq.~(\ref{flux-balance-equation}) implies
\begin{align}
J^{(1)} = J^{(2)} = J, \quad \text{a constant current}, \label{flux-balance-eq-in-steady-state}
\end{align}
which leads to
\begin{equation}
\label{rho1-rho2-soln}
\rho^{(1)} (1 - \rho^{(1)}) = \rho^{(2)} (1 - \rho^{(2)}),
\end{equation}
where \(\rho^{(1)}\) and \(\rho^{(2)}\) denote the steady-state densities on \(T_1\) and \(T_2\), respectively. Thus either $\rho^{(1)} = \rho^{(2)}$, or $\rho^{(1)} + \rho^{(2)} = 1$.

Let us now identify the possible phases in the system. Solving the steady-state current-density relation
\begin{equation}
J = \rho(1-\rho),
\end{equation}
for each TASEP lane yields two solutions for the density:
\begin{equation}
\label{rho_pm}
\rho = \frac{1}{2} \left( 1 \pm \sqrt{1 - 4J} \right) \coloneqq \rho_{\pm}.
\end{equation}
Thus, the steady-state density in each lane can be spatially uniform in one of three possible phases: the low-density (LD) phase where $\rho = \rho_{-} < 1/2$, the high-density (HD) phase where $\rho = \rho_{+} > 1/2$, or the maximal current (MC) phase where \(\rho = 1/2\). Another possibility is a spatially nonuniform density profile that connects LD and HD domains, known as the domain wall (DW) phase or shock phase (SP).

Since the model comprises of two TASEP lanes, each capable of being in one of the four phases, one might expect \(4^2 = 16\) possible phase combinations. However, not all combinations are realized, being not permitted by the dynamics. It turns out that the {\em only} solutions for the stationary densities as admitted by the dynamics are those which give equal stationary densities in both the lanes: $\rho^{(1)} = \rho^{(2)}$ in the bulk, giving LD-LD, HD-HD, MC-MC, and DW-DW phases. The unequal steady state solutions where $\rho^{(1)}\neq \rho^{(2)}$ are absent as a consequence of the boundary conditions controlled by $\alpha_1=\alpha_2$ and $\beta_1=\beta_2$; see also Sec.~\ref{exclusion-of-asym-phases} for more discussions.

\section{MEAN-FIELD PHASE DIAGRAMS OF THE MODEL FOR $0 < \mu \le 1$}
 \label{pd}

 \begin{figure*}[htb]
 \includegraphics[width=\textwidth]{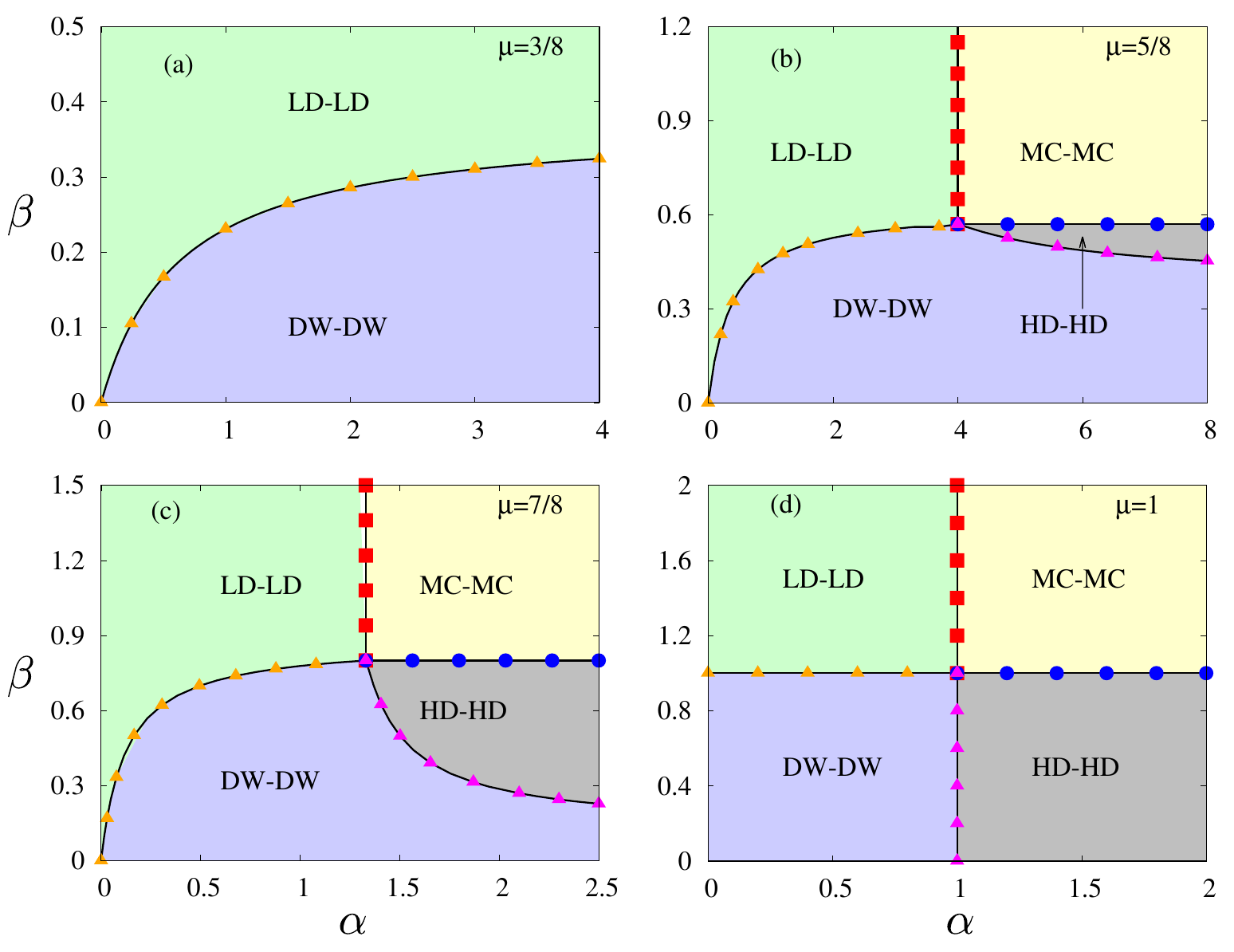}
 \\
\caption{Phase diagrams in the $\alpha$-$\beta$ space with a set of representative values of $0 < \mu \le 1$. Depending on the value of $\mu$, which can take values between 0 and 2, two or four phases are simultaneously found. These are LD-LD, HD-HD, MC-MC, and DW-DW and are represented by green, gray, yellow, and blue regions, respectively, according to MFT. These phases are delineated by black solid lines according to MFT, see Eqs.~(\ref{ldld-and-mcmc-boundary}), (\ref{hdhd-and-mcmc-boundary}), (\ref{ldld-and-dwdw-bundary}), and (\ref{hdhd-and-dwdw-bundary}). MCS confirm the phase boundaries with colored points: red squares (between LD-LD and MC-MC), blue circles (between HD-HD and MC-MC), orange triangles (between LD-LD and DW-DW), and magenta triangles (between HD-HD and DW-DW). System size is $L=1000$ and $2 \times 10^{9}$ Monte Carlo steps are taken. Phase diagrams for $1 < \mu \le 2$ are presented in Fig.~\ref{ph-pd}, exhibiting the particle-hole symmetry of the model.}
\label{phase-diagrams-of-the-model}
\end{figure*}

 Before we analyze the phase diagrams of the present model, it is useful to review of the phase diagram of an open TASEP~\cite{chou-mallick-zia,blythe,tasep-rev}. In an open TASEP, depending on the entry or exit rates ($\alpha_{T}$ and $\beta_{T}$), one finds LD, HD, and MC phases with (constant) steady state densities $\rho_\text{LD}=\alpha_{T}$, $\rho_\text{HD}=1-\beta_{T}$, and $\rho_\text{MC}=1/2$, respectively, in the bulk. Conditions for obtaining these phases are: $\alpha_{T}<\beta_{T}$ and $\alpha_{T}<1/2$ (LD phase), $\beta_{T}<\alpha_{T}$ and $\beta_{T}<1/2$ (HD phase), and $\alpha_{T},\beta_{T}\ge1/2$ (MC phase). In addition to these, a \textit{delocalized} domain wall (DDW) is also formed for $\alpha_{T}=\beta_{T}<1/2$. This DDW is manifested as an inclined line spanning the entire TASEP lane and is a consequence of the particle \textit{nonconserving} dynamics in an open TASEP. Now, transitions between different phases can be identified by equating the particle currents in the associated phases. Thus, the boundaries between either LD or HD and MC phases come out to be straight lines parallel to the $\beta_{T}$ and $\alpha_{T}$ axes respectively, wherein the LD and HD phases are separated by a \textit{coexistence line} (CL) which starts at (0,0) and terminates at (1/2,1/2) in the $\alpha_{T}$-$\beta_{T}$ plane. Considering the bulk density, $\rho$, as the order parameter, one finds the transition from LD to HD phase as discontinuous change in $\rho$, thereby implying it a \textit{discontinuous} phase transition. Contrary to this, the transition from either LD or HD to MC phase is seen to be a continuous change in $\rho$, implying it a \textit{continuous} phase transition.

 We now illustrate the phase diagrams of our model in the plane of control parameters $\alpha$ and $\beta$ for a set of representative values of filling factor $\mu$ (cf. Fig.~\ref{phase-diagrams-of-the-model} and \ref{ph-pd}). We use mean-field theory to obtain the phase diagrams and density profiles, which are corroborated by an extensive Monte Carlo simulation (MCS) studies with random sequential updates. We note that more ordered update schemes, such as parallel (synchronous) updates, have also been widely used in the context of traffic and transport models; see, for instance, Refs.~\cite{rajewsky-update,schreckenberg-traffic}, which may lead to a simpler, deterministic bulk dynamics. We restrict ourselves here to using random sequential updates, which are well-suited to capturing the steady states. MFT and MCS results agree well with each other. In our model, depending on the value of $\mu$, two or four distinct phases appear simultaneously. These are LD-LD, HD-HD, MC-MC and DW-DW phases. Interestingly, the domain wall solutions, which form the DW-DW phase in our model, always emerge in pair for a particular set of values of $\alpha,\,\beta$, and $\mu$ are actually moving density shocks, not pinned to specific locations in the TASEP lanes, having envelops spanning a finite region across the lanes. These are the \textit{delocalized domain walls} (DDW) mentioned above. {For small values of $\mu$, such as $\mu = 3/8$ [Fig.~\ref{phase-diagrams-of-the-model}(a)], only the LD-LD and DW-DW phases are present, separated by a curved phase boundary. This structure persists for all $0 < \mu < 1/2$, with the LD-LD region gradually shrinking as $\mu$ increases. In this regime, the HD-HD and MC-MC phases are absent. As $\mu$ increases beyond $1/2$, the MC-MC and HD-HD phases start to emerge, for instance see the phase diagram at $\mu = 5/8$ [Fig.~\ref{phase-diagrams-of-the-model}(b)]. The LD-LD and HD-HD regions now share boundaries with the MC-MC phase, and these boundaries are straight lines parallel to the $\beta$ and $\alpha$ axes, respectively, independent of the value of $\mu$. This topological structure remains unchanged at $\mu = 7/8$ [Fig.~\ref{phase-diagrams-of-the-model}(c)], where the HD-HD region continues to grow while the LD-LD and DW-DW regions contract. At the half-filling point ($\mu = 1$) [Fig.~\ref{phase-diagrams-of-the-model}(d)], the phase diagram simplifies further. All the phase boundaries become straight lines, aligned with the coordinate axes in the $\alpha$-$\beta$ phase diagram. Moreover, the particle-hole symmetry of the model is used to construct the phase diagrams above half-filling ($\mu > 1$) from their counterparts below half-filling; see Fig.~\ref{ph-pd} in Appendix~\ref{ph-and-pd}. From these, it is evident that as $\mu$ increases, the LD-LD region diminishes while the HD-HD and DW-DW regions expand. For $\mu > 3/2$, both the LD-LD and MC-MC phases vanish, leaving only the HD-HD and DW-DW phases, a trend that continues up to $\mu = \mu_\text{max} = 2$. Thus, for $1/2 < \mu < 3/2$, all four phases — LD-LD, HD-HD, MC-MC, and DW-DW — coexist and meet at a single point known as the \textit{four-phase multicritical point}.}

 \section{STEADY STATE DENSITIES AND PHASE BOUNDARIES}
 \label{steady-state-density}

  In this section, we employ MFT to determine the steady-state bulk densities in the four possible phases of the model — LD-LD, HD-HD, MC-MC, and DW-DW — as well as the phase boundaries that separate them. We also provide a detailed explanation of why asymmetric phases such as LD-HD, LD-DW, HD-DW, and their particle-hole symmetric counterparts do not appear in our model.

 \subsection{LD-LD phase}
 \label{LD-LD-Phase}

 In the LD-LD phase, the stationary densities in both TASEP lanes are equal:
\[
\rho^{(1)} = \rho^{(2)} \equiv \rho_\text{LD-LD}<\frac{1}{2},
\]
where $\rho^{(1)}$ and $\rho^{(2)}$ denote the bulk densities in $T_1$ and $T_2$, respectively. In the LD-LD phase,  $\rho^{(1)}=\alpha_\text{eff}^{(1)}<1/2$ and $\rho^{(2)}=\alpha_\text{eff}^{(2)}<1/2$. Equating the two
\begin{align}
 &\alpha_\text{eff}^{(1)}=\alpha_\text{eff}^{(2)} \nonumber \\
 \implies &\alpha \frac{N_{1}}{L}=\alpha \frac{N_{2}}{L} \nonumber \\
 \implies &N_{1}=N_{2}. \label{ldld-n1-n2}
 \end{align}
PNC relation in this phase reads:
 \begin{align}
  N_{0}=N_{1}+N_{2}+L(\rho^{(1)}+\rho^{(2)})
 \end{align}
Recalling $N_{0}=2\mu L$ and using Eq.~(\ref{ldld-n1-n2}), we obtain $N_{1}$ and $N_{2}$ in terms of the control parameters $\alpha$ and $\mu$:
\begin{equation}
 \label{n1-n2-for-ld-ld-phase}
  \frac{N_{1}}{L}=\frac{\mu}{1+\alpha}=\frac{N_{2}}{L}.
 \end{equation}
Thus, $N_{1} = N_{2}$;  they scale linearly with system size $L$ and filling factor $\mu$. {Since, $0 < N_{1},N_{2} < L$, Eq.~(\ref{n1-n2-for-ld-ld-phase}) yields
\begin{equation}
\label{ldld-ineq-1}
\mu > 0 \quad \text{and} \quad \frac{\mu}{1+\alpha} < 1.
\end{equation}
The effective rates [Eqs.~(\ref{eff-rates-T1})-(\ref{f-and-g})] in the LD-LD phase can now be obtained using Eq.~(\ref{n1-n2-for-ld-ld-phase}) as
\begin{eqnarray}
  &&\alpha_\text{eff}^{(1)}=\alpha_\text{eff}^{(2)}=\frac{\mu \alpha}{1+\alpha}, \label{alphaeff-ldld}\\
  &&\beta_\text{eff}^{(1)}=\beta_\text{eff}^{(2)}=\beta \bigg( 1-\frac{\mu}{1+\alpha} \bigg). \label{betaeff-ldld}
 \end{eqnarray}
 Since both TASEP lanes are in the LD phases, we require $\alpha_\text{eff}^{(i)}<\beta_\text{eff}^{(i)}$ and $\alpha_\text{eff}^{(i)}<1/2$ for $i=1,2$, in analogy with an open boundary TASEP. Substituting Eqs.~(\ref{alphaeff-ldld}) and (\ref{betaeff-ldld}) into these conditions gives:
 \begin{eqnarray}
  &&\frac{\mu}{1+\alpha}<\frac{\beta}{\alpha+\beta}, \label{ldld-ineq-2}\\
  &&\frac{\mu \alpha}{1+\alpha}<\frac{1}{2}. \label{ldld-ineq-3}
 \end{eqnarray}
 Inequalities~(\ref{ldld-ineq-1}), (\ref{ldld-ineq-2}), (\ref{ldld-ineq-3}) together define the LD-LD region in the $\alpha$-$\beta$ plane for a fixed $\mu$, see the LD phase regions in the phase diagrams of Figs.~\ref{phase-diagrams-of-the-model} and \ref{ph-pd}.
}

  Using Eq.~(\ref{alphaeff-ldld}), steady-state densities in both lanes in the LD-LD phase are obtained as:
\begin{equation}
\label{rho-for-ld-ld-phase}
\rho_\text{LD-LD} = \frac{\mu \alpha}{1 + \alpha}.
\end{equation}
As expected, $\rho_\text{LD-LD}$ is independent of the exit rate parameter $\beta$. {For $\mu = \mu_\text{min} = 0$, we have $N_{1} = N_{2} = 0$ [from Eq.~(\ref{n1-n2-for-ld-ld-phase})], and the bulk density is $\rho_\text{LD-LD} = 0$ [from Eq.~(\ref{rho-for-ld-ld-phase})], indicating that both TASEP lanes are empty, as expected.
} The dependence of $\rho_\text{LD-LD}$ on $\alpha$ is shown in Fig.~\ref{rhold-rhohd-vs-a-b}(a), where it can be observed that for large values of $\alpha$, $\rho_\text{LD-LD}$ saturates to $\mu$. 

At this point, we find the range of $\mu$ that permits LD-LD phase in the phase diagram. This follows from the inequalities~\eqref{ldld-ineq-1} and \eqref{ldld-ineq-3}, which respectively provide the lower and upper bounds on $\alpha$. These bounds can be satisfied simultaneously only when the parameter $\mu$ lies in the range
\begin{equation}
\label{ldld-mu-range}
0 < \mu < \frac{3}{2}.
\end{equation}
This is consistent with the Monte Carlo simulation (MCS) results in Figs.~\ref{phase-diagrams-of-the-model} and \ref{ph-pd} where the LD-LD phase appears for $0 < \mu < 3/2$ and is \textit{absent} for \(3/2 < \mu < 2\).

 \subsection{HD-HD phase}
 \label{HD-HD-Phase}

 The HD-HD phase is particle-hole symmetric counterpart to the LD-LD phase. Consequently, by applying the transformations~\eqref{tr-2}--\eqref{tr-3}, the effective entry and exit rates, phase region and steady state density of the HD-HD phase can be directly obtained from the results of the LD-LD phase discussed in the previous section.

Similar to the LD-LD phase, both reservoirs are equally populated in the HD-HD phase (see Appendix~\ref{hdhd-detailed}):
\begin{equation}
\label{n1-n2-for-hd-hd-phase}
\frac{N_1}{L} = \frac{\mu - 1 + \beta}{1 + \beta} = \frac{N_2}{L}.
\end{equation}
The effective entry and exit rates in the HD-HD phase follow directly from the corresponding effective rates of the LD-LD phase, given in Eqs.~\eqref{alphaeff-ldld} and \eqref{betaeff-ldld}, upon applying the particle-hole transformations~\eqref{tr-2} and \eqref{tr-3}:
\begin{eqnarray}
  &&\alpha_\text{eff}^{(1)}=\alpha_\text{eff}^{(2)}=\frac{\alpha(\mu-1+\beta)}{1+\beta}, \label{alphaeff-hdhd}\\
  &&\beta_\text{eff}^{(1)}=\beta_\text{eff}^{(2)}=\frac{(2-\mu)\beta}{1+\beta}. \label{betaeff-hdhd}
\end{eqnarray}
The HD-HD region in the parameter space can similarly be obtained from the LD-LD region, defined by inequalities~\eqref{ldld-ineq-1}, \eqref{ldld-ineq-2} and \eqref{ldld-ineq-3}, using the same set of transformations~\eqref{tr-2} and \eqref{tr-3}. This yields the following conditions:
\begin{equation}
\label{hdhd-ineq-1}
\mu \le 2 \quad \text{and} \quad \frac{2-\mu}{1+\beta} < 1,
\end{equation}
\begin{eqnarray}
  &&\frac{2-\mu}{1+\beta} < \frac{\alpha}{\alpha+\beta}, \label{hdhd-ineq-2}\\
  &&\frac{(2-\mu)\beta}{1+\beta} < \frac{1}{2}. \label{hdhd-ineq-3}
\end{eqnarray}

Inequalities~(\ref{hdhd-ineq-1}), (\ref{hdhd-ineq-2}), (\ref{hdhd-ineq-3}) together define the HD-HD region in the $\alpha$-$\beta$ plane for a fixed $\mu$, see the HD region displayed in the phase diagrams of Figs.~\ref{phase-diagrams-of-the-model} and \ref{ph-pd}.

\begin{figure*}[htb]
 \includegraphics[width=\textwidth]{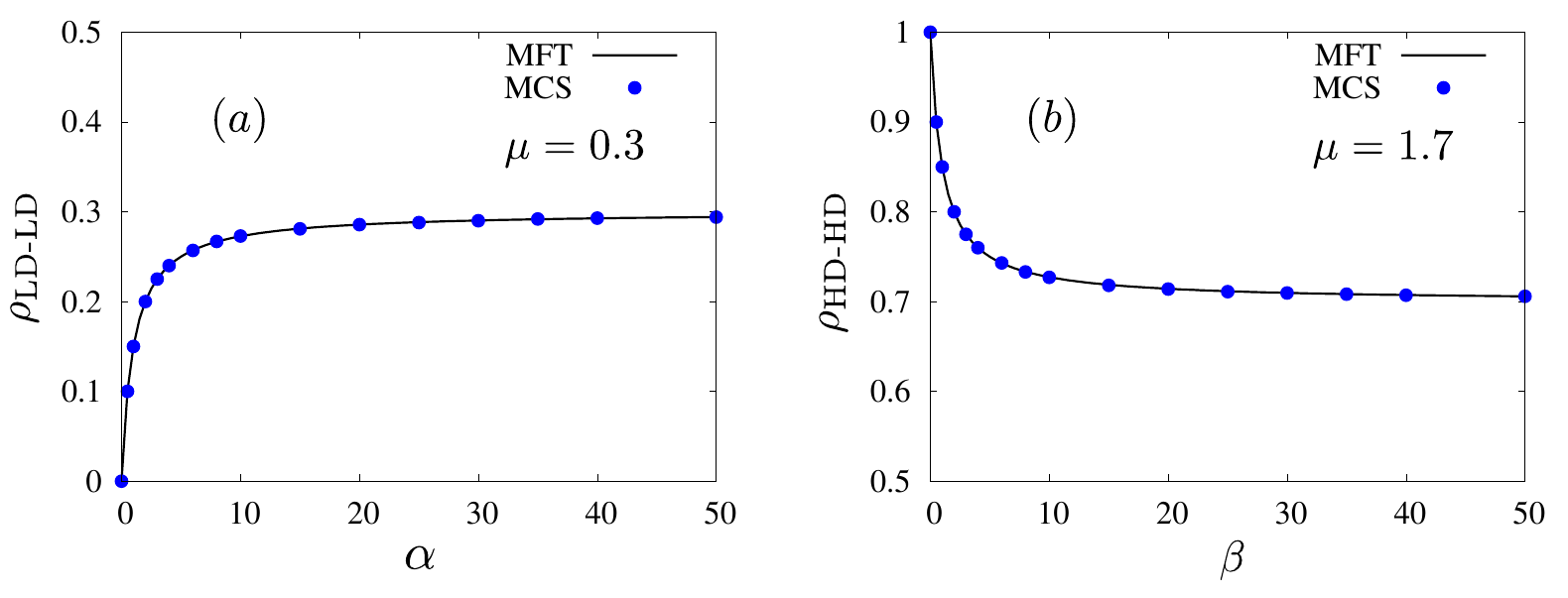}
 \\
\caption{(a) Plot of $\rho_{\text{LD-LD}}$ as a function of $\alpha$ for fixed $\mu=0.3$ [see Eq.~(\ref{rho-for-ld-ld-phase})]. For large $\alpha$, $\rho_{\text{LD-LD}}$ approaches $\mu$. (b) Plot of $\rho_{\text{HD-HD}}$ as a function of $\beta$ for fixed $\mu=1.7$ [see Eq.~(\ref{rho-for-hd-hd-phase})]. For large $\beta$, $\rho_{\text{HD-HD}}$ approaches $\mu - 1$. MFT and MCS results show excellent agreement.}
\label{rhold-rhohd-vs-a-b}
\end{figure*}

The steady-state densities in both lanes in the HD-HD phase can be obtained from Eq.~\eqref{rho-for-ld-ld-phase} using the trasformations~\eqref{tr-1} and \eqref{tr-3}:
\begin{equation}
\label{rho-for-hd-hd-phase}
\rho_\text{HD-HD} = \frac{1 - \beta + \mu \beta}{1 + \beta}.
\end{equation}
As expected, $\rho_\text{HD-HD}$ is independent of the entry rate parameter $\alpha$ and is governed solely by the exit dynamics. {When $\mu = \mu_\text{max} = 2$, we have $N_{1} = N_{2} = L$ [from Eq.~(\ref{n1-n2-for-hd-hd-phase})], and the bulk density is $\rho_\text{HD-HD} = 1$ [from Eq.~(\ref{rho-for-hd-hd-phase})], indicating that both TASEP lanes are fully occupied with $L$ particles in each. Consequently, the system is completely filled with a total particle number $N_{0} = 4L$.
} Moreover, for large values of $\beta$, the density approaches $(\mu - 1)$, consistent with the trend observed in Fig.~\ref{rhold-rhohd-vs-a-b}(b). \par

The range of $\mu$ over which the HD-HD phase exists can be directly obtained from the corresponding range for the LD-LD phase by applying the transformation $\mu \to 2-\mu$; see the transformation (\ref{tr-3}). This yields the following condition for the HD-HD phase:
\begin{equation}
\label{hdhd-mu-range}
\frac{1}{2} < \mu < 2.
\end{equation}

These analytical results are consistent with the MCS results presented in the phase diagrams of Fig.~\ref{phase-diagrams-of-the-model} and Fig.~\ref{ph-pd} where the HD-HD phase is found for $1/2 < \mu < 2$ and is \textit{absent} for \(0 < \mu < \frac{1}{2}\).

 \subsection{MC-MC phase}
 \label{MC-MC-Phase}

 In the MC-MC phase, both lanes maintain a density \(\rho^{(1)}=\rho^{(2)} \equiv \rho_\text{MC-MC} = 1/2\) in the bulk, corresponding to the maximal current \(J_\text{max} = 1/4\). In analogy with an open TASEP, 
\begin{eqnarray}
  &&\alpha_\text{eff}^{(1)} = \alpha \frac{N_1}{L} > \frac{1}{2}, \label{a-1} \\
  &&\beta_\text{eff}^{(1)} = \beta \left(1 - \frac{N_2}{L} \right) > \frac{1}{2}, \label{b-1} \\
  &&\alpha_\text{eff}^{(2)} = \alpha \frac{N_2}{L} > \frac{1}{2}, \label{a-2} \\
  &&\beta_\text{eff}^{(2)} = \beta \left(1 - \frac{N_1}{L} \right) > \frac{1}{2}, \label{b-2}
\end{eqnarray}
in the MC-MC phase.
These give
\begin{equation}
\label{n-lim-mcmc}
\frac{1}{2\alpha} < \frac{N_a}{L} < 1 - \frac{1}{2\beta}, \quad a=1,2, 
\end{equation}
which leads to the constraint
\begin{equation}
\label{a-b-cond-mcmc}
\frac{1}{\alpha} + \frac{1}{\beta} < 2 \quad \Leftrightarrow \quad \alpha + \beta < 2\alpha\beta.
\end{equation}

{The PNC condition in the MC-MC phase reads
\begin{align}
  &N_{0} = 2\mu L = N_1 + N_2 + L(\rho^{(1)} + \rho^{(2)}) \nonumber \\
  \implies &(2\mu - 1) = \frac{N_1}{L} + \frac{N_2}{L}. \label{pnc-mcmc}
\end{align}
Combining the conditions~(\ref{a-1}) and (\ref{a-2}) with Eq.~(\ref{pnc-mcmc}) gives
\begin{equation}
  \label{mcmc-ineq-1}
  \mu > \frac{1}{2}\left(1 + \frac{1}{\alpha}\right).
\end{equation}
Similarly, combining conditions~(\ref{b-1}) and (\ref{b-2}) with Eq.~(\ref{pnc-mcmc}) gives
\begin{equation}
  \label{mcmc-ineq-2}
  \mu < \frac{1}{2}\left(3 - \frac{1}{\beta}\right).
\end{equation}
Inequalities~(\ref{a-b-cond-mcmc}), (\ref{mcmc-ineq-1}), and (\ref{mcmc-ineq-2}) together define the MC-MC region in the $\alpha$–$\beta$ phase diagram for a fixed $\mu$.
}

To find the range of $\mu$ for MC-MC phase existence, consider first the case \(\mu = 1/2 + \epsilon\), where \(0 < \epsilon < 1\). Substituting into inequality~(\ref{mcmc-ineq-1}) yields
\begin{equation}
\label{mcmc-ineq-1-modified}
\frac{1}{\alpha} < 2\epsilon,
\end{equation}
and inequality~(\ref{mcmc-ineq-2}) becomes
\begin{equation}
\label{mcmc-ineq-2-modified}
\frac{1}{\beta} < 2(1 - \epsilon).
\end{equation}
These conditions can be satisfied for appropriate values of \(\alpha\) and \(\beta\), consistent with~(\ref{a-b-cond-mcmc}), confirming that the MC-MC phase is allowed when \(1/2 < \mu < 3/2\).
Now, consider \(\mu = 1/2 - \epsilon\) with \(0 < \epsilon < 1/2\). Then inequality~(\ref{mcmc-ineq-1}) becomes:
\begin{equation}
\label{mcmc-ineq-1-modified-2}
\frac{1}{\alpha} < -2\epsilon,
\end{equation}
which is impossible since \(\alpha > 0\) and the right-hand side is negative. Hence, MC-MC cannot occur in this regime.
Similarly, for \(\mu = 3/2 + \epsilon\) with \(0 < \epsilon < 1/2\), inequality~(\ref{mcmc-ineq-2}) becomes:
\begin{equation}
\label{mcmc-ineq-2-modified-2}
\frac{1}{\beta} < -2\epsilon,
\end{equation}
which again cannot be satisfied for \(\beta > 0\). This rules out the MC-MC phase for \(\mu > 3/2\).
Therefore, the MC-MC phase is restricted to the interval \( 1/2 < \mu < 3/2 \), in full agreement with the MCS observations in Figs.~\ref{phase-diagrams-of-the-model} and~\ref{ph-pd}.

This result can also be understood from the particle number conservation. In the MC-MC phase, each lane contains \(L/2\) particles, so the total number of particles in the lanes is \(L\). Using the relation \(N_0 = 2\mu L\), particle number conservation implies:
\[
N_1 + N_2 = N_0 - L = (2\mu - 1)L.
\]
To ensure \(N_1 + N_2 > 0\), it follows that \(\mu > 1/2\). Furthermore, since \(N_1, N_2 \le L\), the total number of particles in the system must satisfy \(N_0 \leq 3L\) in the MC-MC phase, which in turn implies \(\mu < 3/2\).

\begin{figure}[htb]
 \includegraphics[width=\columnwidth]{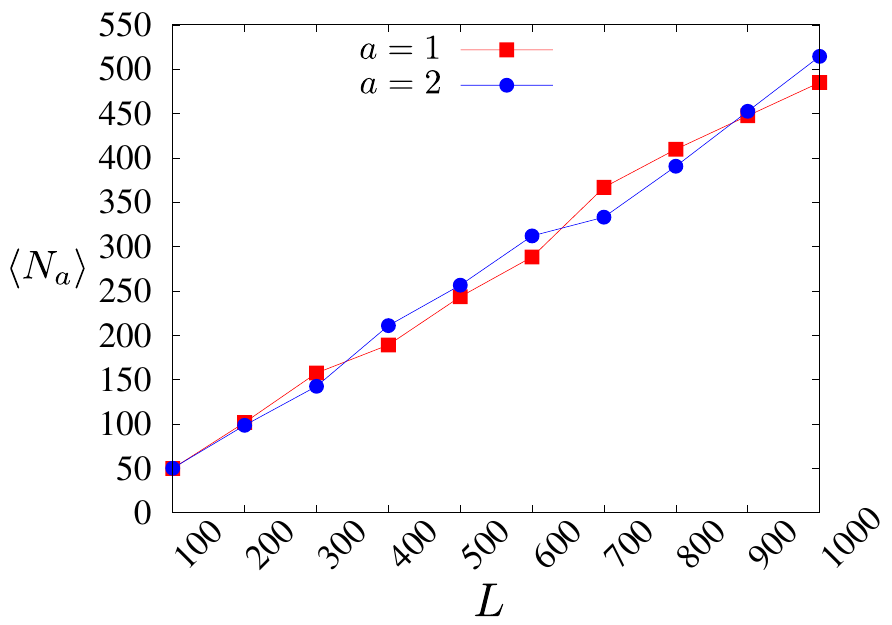}
 \caption{Steady state time-averaged reservoir populations $\langle N_1\rangle,\langle N_2\rangle$ when the TASEP lanes are in their MC-MC phase. Within error bars, $\langle N_1\rangle =\langle N_2\rangle$ is found.}\label{mc-mc-res}
\end{figure}

Notice that in the MC-MC phase, the MFT gives only ranges of $N_1$ and $N_2$; see Eq.~\eqref{n-lim-mcmc} above. We examine this further numerically and solve for $N_1,\,N_2$ numerically in the steady state, averaged over large time. This gives us $\langle N_1\rangle$ and $\langle N_2\rangle$, where $\langle . \rangle$ implies temporal averages in the steady states. Our numerical results are shown in Fig.~\ref{mc-mc-res}, which clearly establishes $N_1=N_2$ within error bars. 

 \subsection{DW-DW phase}
 \label{dwdw-phase}

 We now investigate the nonuniform steady states in which domain walls are formed connecting the LD and HD phases. Assuming both the TASEP lanes to have one DW,  for a particular choice of $\alpha$ and $\beta$, let the DWs formed in $T_{i}$ be located at $x_{w}^{(i)}$, $i=1,2$. In MFT, these DWs are given by:
\begin{equation}
  \label{dw-theta}
  \rho^{(i)}(x)=\rho_{\text{LD}}^{(i)}+\Theta(x-x_{w}^{(i)})(\rho_{\text{HD}}^{(i)}-\rho_{\text{LD}}^{(i)}),
 \end{equation}
 where
$\Theta(x)$ is the Heaviside step function defined as $\Theta(x)=1(0)$ for $x>(<)0$. Similar to an open TASEP, one has $\rho_{\text{LD}}^{(i)}=\alpha_\text{eff}^{(i)}$ and $\rho_{\text{HD}}^{(i)}=1-\beta_\text{eff}^{(i)}$ in the DW-DW phase of the present model. Now, current conservation in the DW-DW phase implies $\rho_{\text{LD}}^{(i)}+\rho_{\text{HD}}^{(i)}=1$ which, along with the solution $\rho^{(1)}=\rho^{(2)}$, infers $\alpha_\text{eff}^{(1)}=\alpha_\text{eff}^{(2)}=\beta_\text{eff}^{(1)}=\beta_\text{eff}^{(2)}$. This in turn gives
\begin{equation}
 \label{n1-n2-for-dwdw-phase}
  \frac{N_{1}}{L}=\frac{\beta}{\alpha+\beta}=\frac{N_{2}}{L},
 \end{equation}
 i.e., $N_1=N_2$
 in the DW-DW phase. Thus, effective entry and exit rates [see Eqs.~(\ref{eff-rates-T1})-(\ref{f-and-g})] in both lanes are obtained as
\begin{equation}
  \alpha_\text{eff}^{(i)}=\beta_\text{eff}^{(i)}=\frac{\alpha \beta}{\alpha+\beta}. \label{alphaeff_betaeff_equals}
 \end{equation}
Consequently, the steady state bulk densities read
\begin{eqnarray}
  &&\rho_{\text{LD}}^{(i)}=\frac{\alpha\beta}{\alpha+\beta}=1-\rho_{\text{HD}}^{(i)}, \label{rholdhd-dwdw-phase}
 \end{eqnarray}
Analogous to the open TASEP, we must have $\alpha_\text{eff}^{(i)}=\beta_\text{eff}^{(i)}<1/2$ in the DW-DW phase of our model. Hence the condition
\begin{equation}
  \frac{1}{\alpha} + \frac{1}{\beta} > 2 \quad \Leftrightarrow \quad \alpha + \beta > 2\alpha\beta. \label{dwdw-region}
 \end{equation}
must be met within the DW-DW region in $\alpha$-$\beta$ phase diagram.

 Next we  determine the locations, $x_{w}^{(1)}$ and $x_{w}^{(2)}$, of the DWs formed in $T_{1}$ and $T_{2}$. The PNC relation in the DW-DW phase translates:
\begin{equation}
2\mu L = N_{1} + N_{2} + L\int_{0}^{1} dx \big[\rho^{(1)}(x) + \rho^{(2)}(x)\big], \label{pnc-relation}
\end{equation}
Plugging the expressions of $N_{1}$ and $N_{2}$ from Eq.~(\ref{n1-n2-for-dwdw-phase}) and densities from Eqs.~(\ref{dw-theta}) and~(\ref{rholdhd-dwdw-phase}) into (\ref{pnc-relation}), we obtain
\begin{equation}
\label{xw1-xw2-eq}
2\left(\mu-1-\frac{\beta}{\alpha+\beta}+\frac{\alpha\beta}{\alpha+\beta}\right) = (x_{w}^{(1)}+x_{w}^{(2)})\left(\frac{2\alpha\beta}{\alpha+\beta}-1\right).
\end{equation}
This equation cannot be solved to determine both $x_{w}^{(1)}$ and $x_{w}^{(2)}$ uniquely keeping the parameters $\alpha,\,\beta$ and $\mu$ fixed, thereupon, indicating the domain walls to be \textit{delocalized} in space. Physically,
shifting a DW in any direction in one lane is compensated by an opposite shift of the DW in another lane, keeping PNC unaffected. Thus all such pairs of $(x_w^{(1)},x_w^{(2)})$ give valid DW positions. The inhererent stochasticity of the dynamics ensures that eventually the system visits all such solutions, displaying only the envelop of all the DWs under long time averaging. Since each channel is uniform having unit hopping rate in the bulk, the probability of finding a DW in a given channel is a constant. This in turn gives that the envelop of such a DW as an inclined lines, i.e., a DDW profile, spanning either the full length of the each lane, or only partially. In case of a partially spanning DDW, the fraction spanned by a DDW in each of the lanes is equal. This is a consequence of PNC. 
Thus in the DW-DW phase, the domain walls in both lanes become delocalized due to the freedom in their relative positions, constrained only by the total particle number conservation, see Eq.~(\ref{xw1-xw2-eq}), see also Ref.~\cite{sm-inhomogeneous-tasep}.

%This is consistent with the particle-hole symmetry of the model.

 One can find the DW height, $\mathcal{H}^{(i)}$, of lanes $T_{i}$, $i=1,2$ in terms of the control parameters as:
\begin{equation}
 \label{height-dwdw-phase}
  \mathcal{H}^{(i)}=\rho_{\text{HD}}^{(i)}-\rho_{\text{LD}}^{(i)}=1-\frac{2\alpha\beta}{\alpha+\beta}.
 \end{equation}
 Note that \(\mathcal{H}^{(i)}\) is independent of \(\mu\), although the positions of the DWs depend on \(\mu\) via Eq.~(\ref{xw1-xw2-eq}). As \(\mu\) increases, additional particles accumulate in the HD regions, pushing the domain walls toward the entry ends.% \textbf{Is the plot $\mathcal{H}$ vs $\mu$ needed?}.{\cred maybe we add a plot}

%%%%%%%%%%%%%%%%%%%
To find the permissible range of \(\mu\) for the DW-DW phase, we evaluate Eq.~(\ref{xw1-xw2-eq}) at the two extreme limits of the domain wall positions. First, consider the case where the domain walls are located at the exit end of both lanes, i.e., \(x_{w}^{(1)} = x_{w}^{(2)} = 1\). In this configuration, the LD domains extend across the entire length of each lane. Conversely, when the domain walls are located at the entry end, i.e., \(x_{w}^{(1)} = x_{w}^{(2)} = 0\), both lanes are entirely in the HD phase. Therefore, \(0 \le (x_{w}^{(1)} + x_{w}^{(2)}) \le 2\) depending on the location of the DWs. Substituting this into Eq.~(\ref{xw1-xw2-eq}), we obtain the following condition on \(\mu\) in the DW-DW regime where \(\alpha+\beta>2\alpha \beta\) (see (\ref{dwdw-region})):
\begin{equation}
\label{dwdw-ineq-1}
\frac{\beta (1+\alpha)}{\alpha+\beta} < \mu < \bigg[1+\frac{\beta (1-\alpha)}{\alpha+\beta}\bigg].
\end{equation}
Inequalities~(\ref{dwdw-region}) and (\ref{dwdw-ineq-1}) together define the DW-DW region in the $\alpha$-$\beta$ phase diagram for any $0 \le \mu \le 2$. This concludes the DW-DW phase is always present in our model, see Figs.~\ref{phase-diagrams-of-the-model}(a)-\ref{phase-diagrams-of-the-model}(d) and Figs.~\ref{ph-pd}(a), \ref{ph-pd}(b).
%%%%%%%%%%%%%%%%%%%%%%%

\begin{figure*}[htb]
\begin{minipage}[b]{0.33\linewidth}
  \centering
  \includegraphics[width=\linewidth]{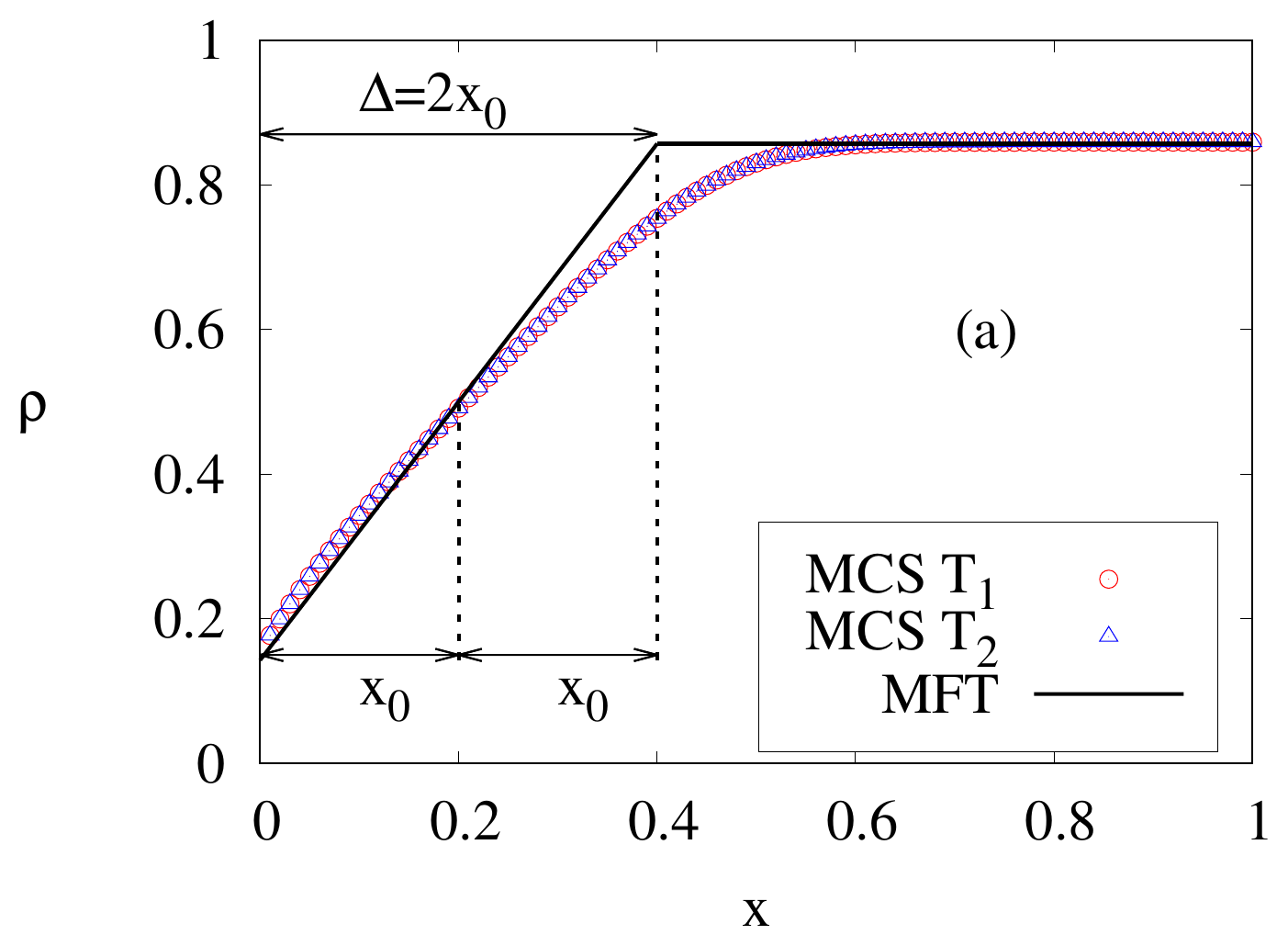}
  %\caption{(a) $x_{0}<\frac{1}{2}$}
\end{minipage}%
\begin{minipage}[b]{0.33\linewidth}
  \centering
  \includegraphics[width=\linewidth]{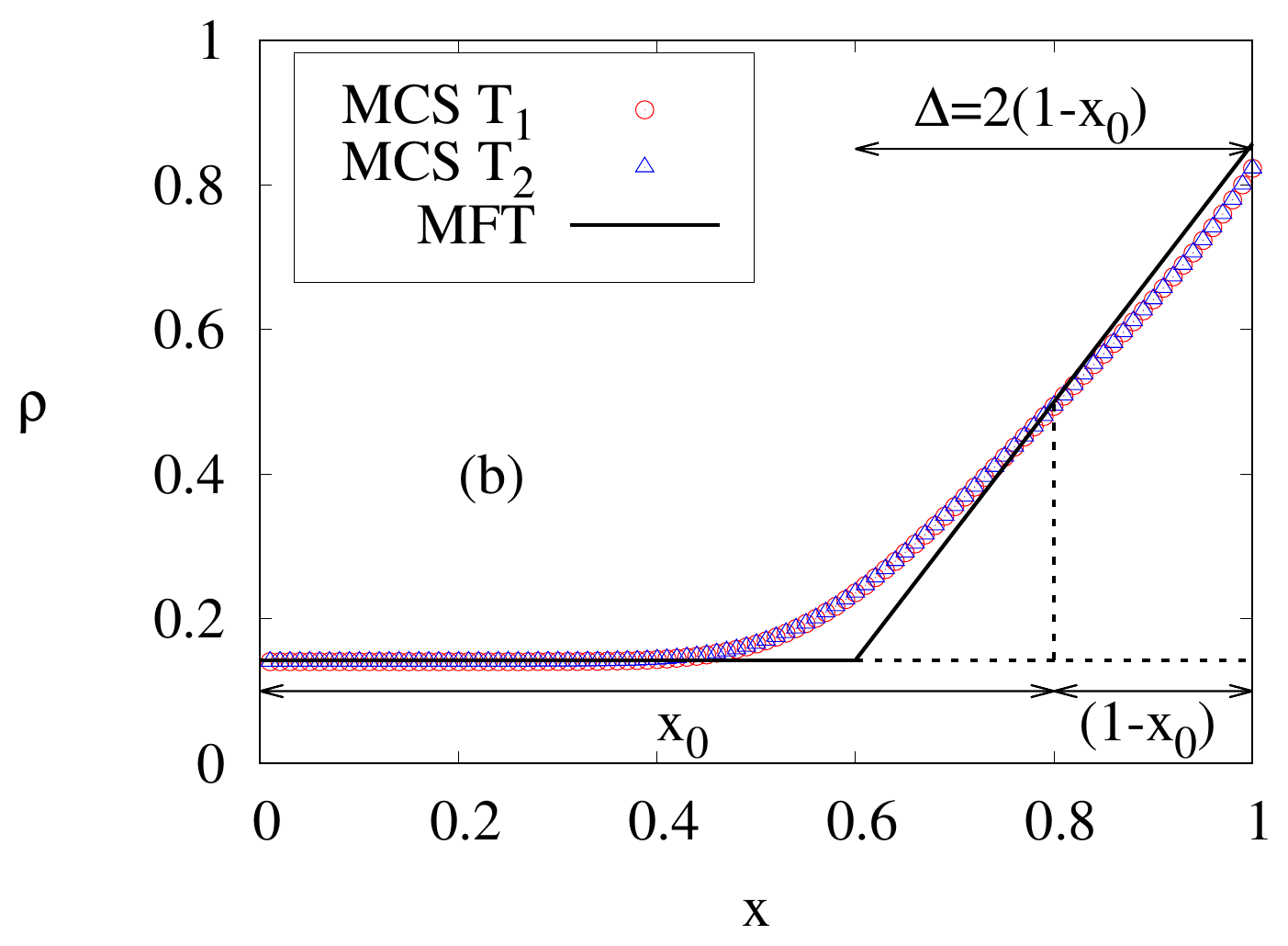}
  %\caption{(b) $x_{0}>\frac{1}{2}$}
\end{minipage}%
\begin{minipage}[b]{0.33\linewidth}
  \centering
  \includegraphics[width=\linewidth]{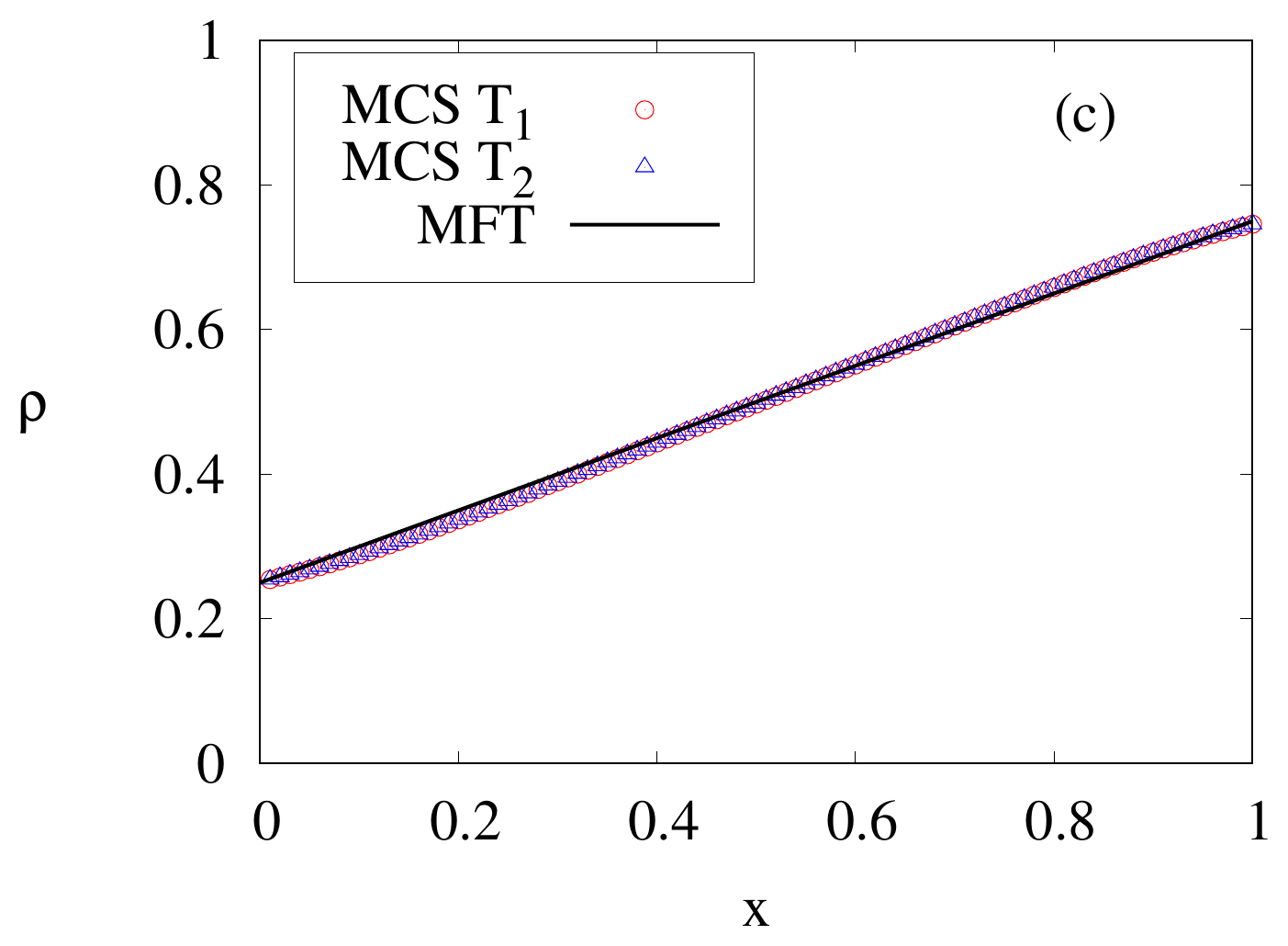}
  %\caption{(c) $x_{0}=\frac{1}{2}$}
\end{minipage}
\caption{Steady state density profiles in both lanes $T_{1}$ and $T_{2}$ are presented in the DW-DW phase. These are delocalized domain walls (DDW). System size is $L=1000$ and $2 \times 10^{9}$ Monte Carlo steps are taken. (a) Partially delocalized with average position $x_{0}<1/2$, span $\Delta=2x_{0}$, and parameter values: $\alpha=0.5$, $\beta=0.2$, and $\mu=1$; (b) Partially delocalized: $x_{0}>1/2$, $\Delta=2(1-x_{0})$, and parameter values: $\alpha=0.2$, $\beta=0.5$, and $\mu=1$; and (c) Fully delocalized: $x_{0}=1/2$, $\Delta=1$, parameter values: $\alpha=\beta=0.5$, and $\mu=1$. MFT predictions and MCS outcomes match well with each other.}\label{ddw-span}
\end{figure*}

%%%%%%%%%%%%%%%%%%%%%%
We now examine whether the DDWs are completely or partially delocalized. The condition $\rho^{(1)} = \rho^{(2)}$ along with PNC, implies that the time-averaged positions of both domain walls satisfy $\langle x_{w}^{(1)} \rangle = \langle x_{w}^{(2)} \rangle = x_{0}$ in the steady state, where $\langle \cdot \rangle$ denotes time-average over steady-state realizations. In a closed system with biased hopping, particles tend to accumulate behind local inhomogeneities (in this case, the reservoirs). Assuming linear profiles for the DDWs, the average domain wall position $x_{0}$ can be determined from PNC as:
\begin{align}
2\mu &= \frac{N_{1}}{L} + \frac{N_{2}}{L}
+ \int_{0}^{x_{0}} \left( \rho_{\text{LD}}^{(1)} + \rho_{\text{LD}}^{(2)} \right)\, dx \notag \\
&\quad + \int_{x_{0}}^{1} \left( \rho_{\text{HD}}^{(1)} + \rho_{\text{HD}}^{(2)} \right)\, dx. \label{pnc-x0-2}
\end{align}
Substituting the expressions from Eqs.~(\ref{n1-n2-for-dwdw-phase}) and~(\ref{rholdhd-dwdw-phase}) into Eq.~(\ref{pnc-x0-2}), we obtain the average wall position in terms of model parameters:
\begin{equation}
\label{x0}
x_{0} = \frac{\alpha + 2\beta - \alpha\beta - \mu(\alpha + \beta)}{\alpha + \beta - 2\alpha\beta}.
\end{equation}
Equation~(\ref{x0}) shows, for fixed values of $\alpha$ and $\beta$, $x_0$ decreases with increasing $\mu$, indicating that the domain walls shift toward the entry-end of the respective lane. Next, we determine the span ($\Delta$) of the DDW, defined as the spatial extent over which the time-averaged DDW density profile (appearing as an inclined line) persists. Table~\ref{tab-ddw-span} analyzes the DDWs using a geometric approach, summarizing the span $\Delta$ corresponding to different values of $x_0$ and the associated conditions on $\mu$. The geometric constructions of DDWs, illustrated in Fig.~\ref{ddw-span}, show excellent agreement with MCS results.
\begin{table}[h!]
\centering
\small % Reduce font size to help with fitting
\renewcommand{\arraystretch}{1.8} % Increased row spacing
\setlength{\tabcolsep}{10pt} % Increased column spacing

\caption{Dependence of the DDW span $\Delta$ on $x_0$ and $\mu$}
\label{tab-ddw-span}

\begin{tabular}{|c|c|c|}
\hline
Condition on $x_0$ & DDW Span $\Delta$ & Condition on $\mu$ \\
\hline
$x_0 < \tfrac{1}{2}$ & $\Delta = 2x_0$ & $\mu > \dfrac{\alpha + 3\beta}{2(\alpha + \beta)}$ \\
\hline
$x_0 = \tfrac{1}{2}$ & $\Delta = 1$ & $\mu = \dfrac{\alpha + 3\beta}{2(\alpha + \beta)}$ \\
\hline
$x_0 > \tfrac{1}{2}$ & $\Delta = 2(1 - x_0)$ & $\mu < \dfrac{\alpha + 3\beta}{2(\alpha + \beta)}$ \\
\hline
\end{tabular}
\end{table}

%Moreover, in the DW-DW phase, the domain walls in both lanes become delocalized due to the freedom in their relative positions, constrained only by the total particle number conservation, see Eq.~(\ref{xw1-xw2-eq}). The detailed behavior -- including whether the delocalization is full or partial -- is discussed in~\ref{dwdw-phase}). 

%{\bf DDW theory?}

%%%%%%%%%%%%%%%%%%%%

\begin{figure*}[htb]
 \includegraphics[width=\textwidth]{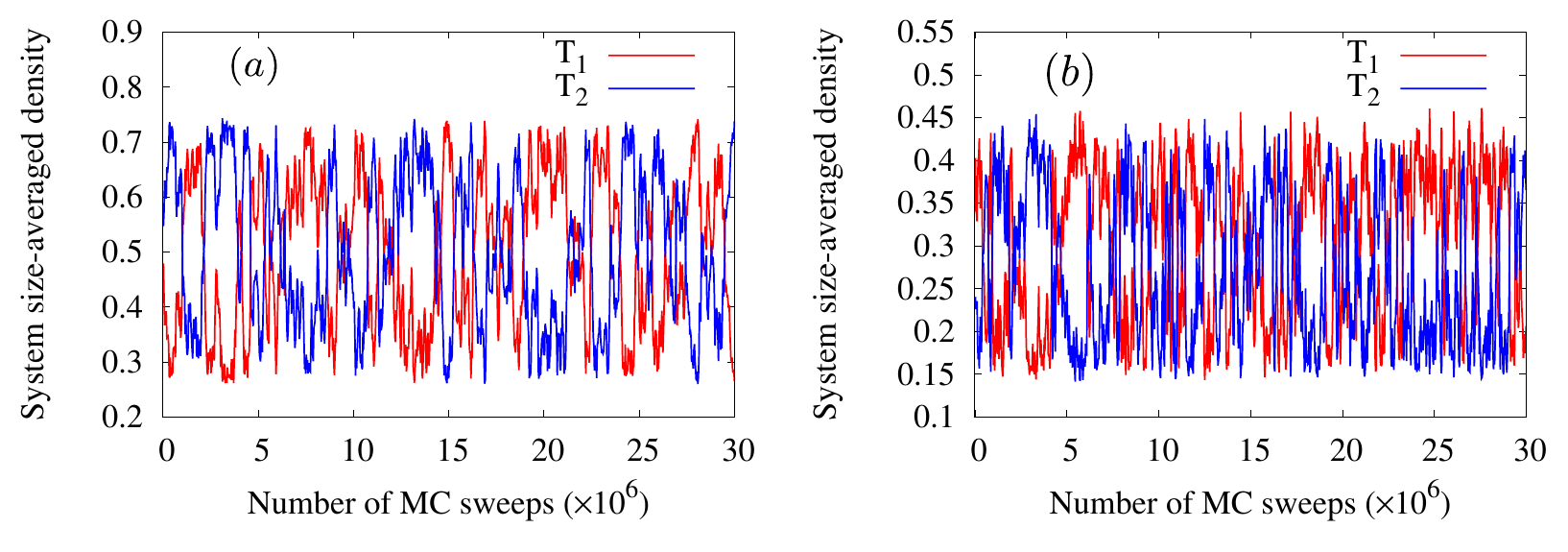}
 \caption{Plots showing the variation of system size-averaged density $\bar \rho_i$ ($i=1,2$ for the two TASEP lanes) with Monte Carlo steps for (a) fully delocalized domain walls with parameters $\alpha=\beta=0.5, \mu=1$ and (b) partially delocalized domain walls with parameters $\alpha=0.2, \beta=0.5, \mu=1$. System size is $L=1000$ and $3\times10^{7}$ MC steps are taken. This shows that fluctuations in stationary densities survive even at thermodynamic limit in the DW-DW phase of our model.}
\label{coarse-den-vs-time}
\end{figure*}

 To explore the extent of number fluctuations in the DW-DW phase, going beyond the MFT, we study the model numerically by MCS. We have noted that the DDWs fluctuate over a distance that scales with $L$. Therefore, the fluctuations in the total particle numbers in the TASEP lanes also scale with $L$ in the DW-DW phase. See Fig.~\ref{coarse-den-vs-time} for plots of the densities in the TASEP lanes, averaged over entire TASEP length, i.e., $L$ as functions of time, measured by the number of Monte-Carlo steps. The plots reveal {\em complementary} nature of the fluctuating densities in the two TASEP lanes.

\begin{figure*}[htb]
 \includegraphics[width=\textwidth]{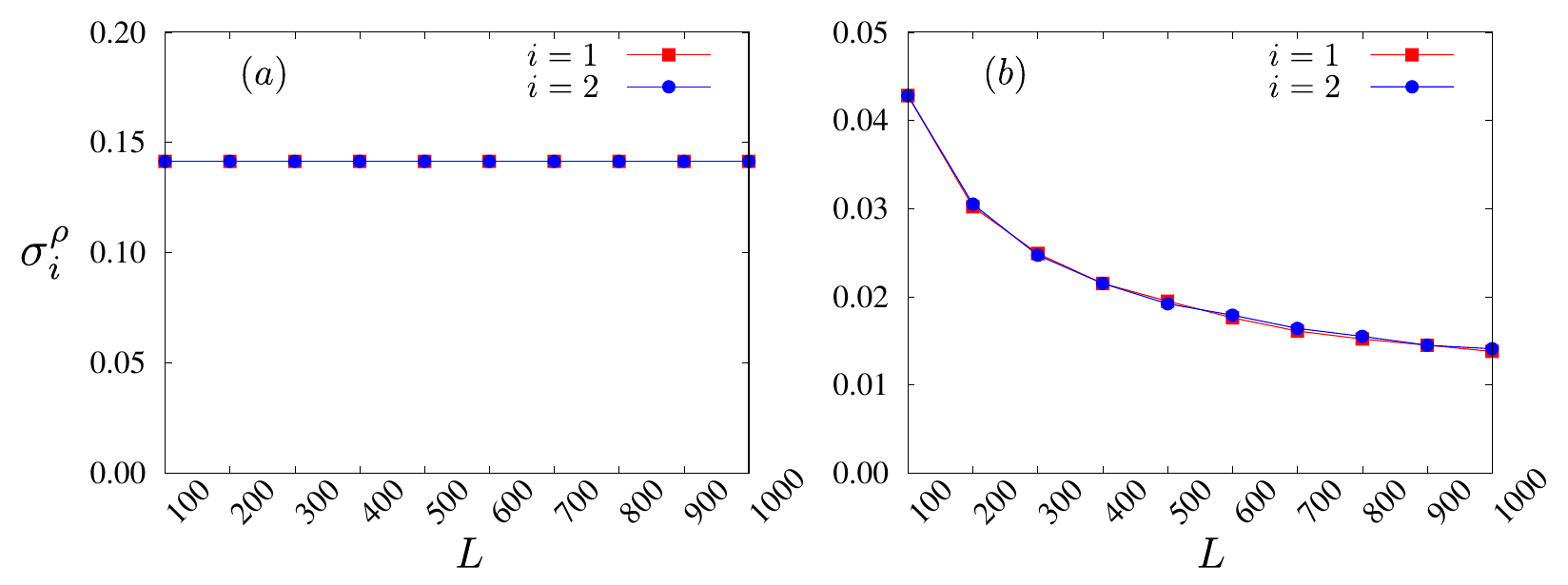}
 \caption{Standard deviations $\sigma^\rho_i$ [see Eq.~(\ref{standard-dev_of_den})] of the system size-averaged densities $\bar\rho_i$ ($i=1,2$ for TASEP lanes $T_{1}$ and $T_{2}$) as functions of system size $L$: (a) for a fully delocalized domain wall with $\alpha=\beta=0.5$ and $\mu=1$, where $\sigma^\rho_i \sim \mathcal{O}(1)$, and (b) for the LD--LD phase with $\alpha=0.5$, $\beta=1.5$, and $\mu=1$, where $\sigma^\rho_i$ scales as $\mathcal{O}(1/L)$, thereby vanishing in the thermodynamic limit.
}
\label{variance-of-den-vs-L}
\end{figure*}

\begin{figure*}[htb]
 \includegraphics[width=\textwidth]{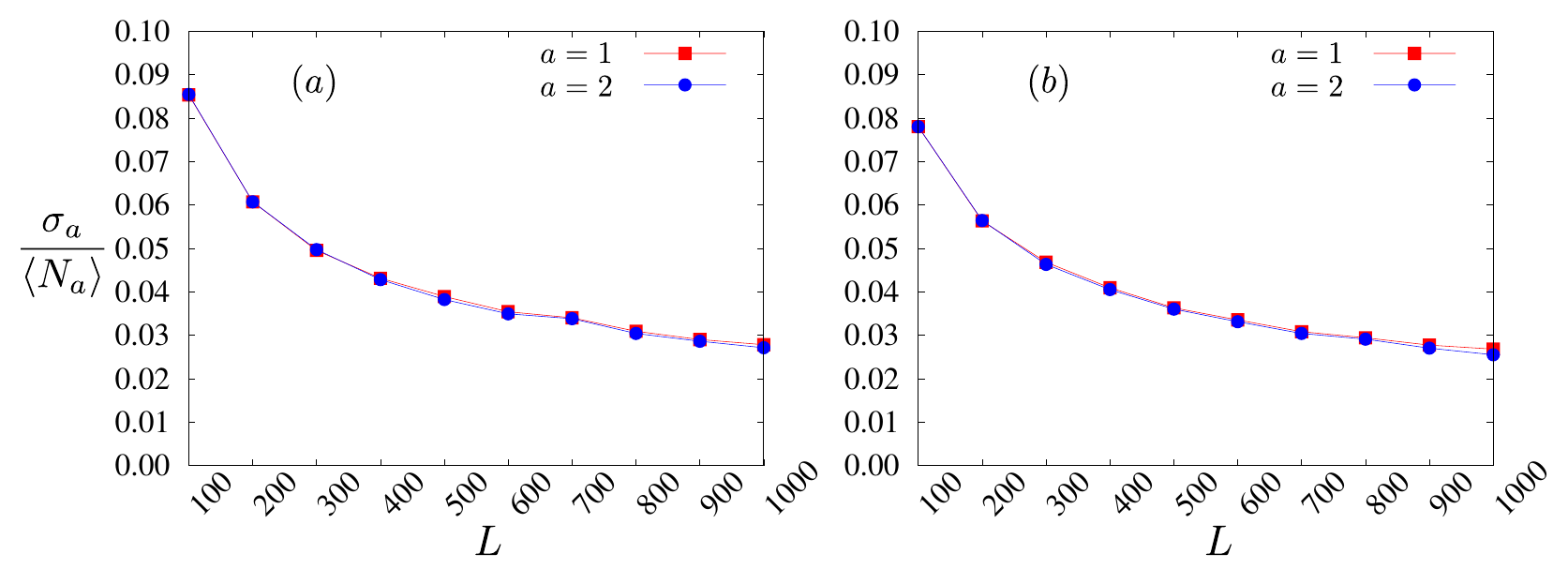}
 \caption{{Plots of the relative fluctuation, $\sigma_{a}/\langle N_{a} \rangle$, of the reservoir populations $N_{a}$ ($a=1,2$ for reservoirs $R_{1}$ and $R_{2}$) as a function of system size $L$. $\sigma_{a} = \sqrt{\langle N_{a}^{2} \rangle - \langle N_{a} \rangle^{2}}$ 
is the standard deviation of $N_{a}$, and $\langle . \rangle$ denotes the time-average. (a) DW--DW phase with complete delocalization for $\alpha = \beta = 0.5$ and $\mu = 1$. (b) LD--LD phase for $\alpha = 0.5$, $\beta = 1.5$, and $\mu = 1$. In both cases, the relative fluctuation decreases monotonically with increasing $L$.}}
\label{n-fluc-res-dwdw-ldld}
\end{figure*}

 To study these density fluctuations quantitatively, we compute the standard deviations $\sigma_i^{\rho}$ of the system size-averaged densities $\bar\rho_i$ ($i = 1, 2$ for the two TASEP lanes), obtained from different Monte Carlo steps. They are defined as  
\begin{equation}
\label{standard-dev_of_den}
\sigma_i^{\rho} = \sqrt{\langle \bar\rho_i^{2} \rangle - \langle \bar\rho_i \rangle^{2}}, \quad i = 1,2, 
\end{equation}
where
\begin{equation}
\bar\rho_i = (\rho_{\text{HD}}^{(i)} - \rho_{\text{LD}}^{(i)})\,\Delta + \rho_{\text{LD}}^{(i)}
\end{equation}
in the DW-DW phase,
$\rho_\text{LD}^{(i)}$ and $\rho_\text{HD}^{(i)}$ are the stationary densities in the low- and high-density domains of a DW obtained in Eq.~(\ref{rholdhd-dwdw-phase}). The DDW span $\Delta$ is same (as found in MFT above, see Table~\ref{tab-ddw-span}) in both the TASEP lanes. We have calculated $\sigma^\rho_i$ for various $L$ from our MCS studies, and found them to be independent of $L$ in the DW-DW phase for a range of $L$ used in our simulations; see Fig.~\ref{variance-of-den-vs-L}(a). Since both $\bar \rho_1,\bar\rho_2$ are ${\mathcal O(1)}$ quantities independent of $L$, this strongly suggests that the relative fluctuations in $ \rho_1,\,\rho_2$ about their mean values $\bar\rho_1,\,\bar\rho_2$ {\em do not vanish} in the DW-DW phase in the thermodynamic limit. This is in contrast to the behavior of $\sigma^\rho_i$ in the LD-LD phase; see Fig.~\ref{variance-of-den-vs-L}(b) with similar behavior expected in the HD-HD and MC-MC phases (not shown). Thus the relative fluctuations in the TASEP densities vanish in the thermodynamic limit in the LD-LD, HD-HD, and MC-MC phases. A pertinent question is whether the reservoir populations $N_1,\,N_2$ {\em also} display large fluctuations  in the thermodynamic limit. We study this question numerically and the results are shown in Fig.~\ref{n-fluc-res-dwdw-ldld}(a) for DW-DW phase and in Fig.~\ref{n-fluc-res-dwdw-ldld}(b) for LD-LD phase, which shows a monotonically decaying trend with a rising $L$, strongly suggesting  the relative fluctuations in the reservoir populations vanish in the thermodynamic limit. This together with the overall particle number conservation explains the observed complementary nature of the fluctuations in the two TASEP densities found in Fig.~\ref{coarse-den-vs-time}. We thus conclude that in the DW-DW phase while the TASEP densities exhibit large fluctuations that survive in the thermodynamic limit, the relative fluctuations in the reservoir populations vanish in the thermodynamic limit. 

%%%%%%%%%%%%%%%%%%%%%%%%

 \subsection{Absence of the asymmetric phases}
 \label{exclusion-of-asym-phases}

We now show that this model with $\alpha_1 = \alpha_2 = \alpha$ and $\beta_1 = \beta_2 = \beta$ does not admit any asymmetric phases such as LD-HD, LD-DW, or HD-DW and their particle-hole symmetric counterparts. % in the $\alpha$-$\beta$ phase diagrams shown in Figs.~\ref{phase-diagrams-of-the-model} and \ref{ph-pd} for the symmetric case with $\alpha_1 = \alpha_2 = \alpha$ and $\beta_1 = \beta_2 = \beta$. In this section, 
We provide a detailed argument for the absence of these phases.

\begin{itemize}
 \item[] \textit{Absence of LD-HD and HD-LD phases:}
 Suppose an LD-HD phase exists where $T_1$ is in the LD phase and $T_2$ in the HD phase, with bulk densities $\rho^{(1)}=\alpha_\text{eff}^{(1)}=\alpha N_{1}/L$ and $\rho^{(2)}=1-\beta_\text{eff}^{(2)}=1-\beta (1-N_{1}/L)$. Current conservation implies:
 \begin{align}
  \rho^{(1)} + \rho^{(2)} &= 1 \nonumber \\
  \Rightarrow \alpha_\text{eff}^{(1)} &= \beta_\text{eff}^{(2)} \nonumber \\
  \Rightarrow \frac{N_1}{L} &= \frac{\beta}{\alpha + \beta}. \label{ldhd-n1}
 \end{align}
 The corresponding value of $N_2$ can be obtained using PNC and Eq.~(\ref{ldhd-n1}), giving
 \begin{equation}
  \frac{N_2}{L} = 2\mu - 1 - \frac{\beta}{\alpha + \beta}. \label{ldhd-n2}
 \end{equation}
 The effective entry and exit rates then read
 \begin{align}
  \alpha_\text{eff}^{(1)} &= \beta_\text{eff}^{(2)} = \frac{\alpha \beta}{\alpha + \beta}, \label{alphaeff1-betaeff2-ldhd} \\
  \alpha_\text{eff}^{(2)} &= \alpha \left(2\mu - 1 - \frac{\beta}{\alpha + \beta}\right), \label{alphaeff2-ldhd} \\
  \beta_\text{eff}^{(1)} &= \beta \left(2 - 2\mu + \frac{\beta}{\alpha + \beta}\right). \label{betaeff1-ldhd}
 \end{align}

 Since $T_1$ is in the LD phase, we must have $\alpha_\text{eff}^{(1)} < \beta_\text{eff}^{(1)}$, leading to
 \begin{equation}
  \mu < \frac{\alpha + 3\beta}{2(\alpha + \beta)}. \label{ldhd-alphaeff1_less-betaeff1}
 \end{equation}
 Similarly, for $T_2$ to be in the HD phase, the condition $\beta_\text{eff}^{(2)} < \alpha_\text{eff}^{(2)}$ must hold, which implies
 \begin{equation}
  \mu > \frac{\alpha + 3\beta}{2(\alpha + \beta)}. \label{ldhd-betaeff2-less-alphaeff2}
 \end{equation}
 The two conditions in Eqs.~(\ref{ldhd-alphaeff1_less-betaeff1}) and (\ref{ldhd-betaeff2-less-alphaeff2}) cannot be simultaneously satisfied, thus ruling out the possibility of an LD-HD phase. A similar argument applies to discard the occurrence of the HD-LD phase, which is the particle-hole symmetric counterpart of LD-HD.

 \item[] \textit{Absence of LD-DW, DW-LD, HD-DW, and DW-HD phases:}
 Let us start with the LD-DW phase, where $T_1$ is in the LD phase and $T_2$ in the DW phase. The bulk densities are given by: $\rho^{(1)}=\alpha_\text{eff}^{(1)}=\alpha N_{1}/L$ and $\rho^{(2)}(x)=\rho^{(2)}_\text{LD}+\Theta(x-x_{w})(\rho^{(2)}_\text{HD}-\rho^{(2)}_\text{LD})$, where $\rho^{(2)}_\text{LD}$ and $\rho^{(2)}_\text{HD}$ are the LD and HD densities of $T_2$, and $x_w$ denotes the domain wall location.

From current conservation, we must have:
\begin{align}
\rho^{(1)} &= \rho^{(2)}_\text{LD} \nonumber \\
\Rightarrow \alpha_\text{eff}^{(1)} &= \alpha_\text{eff}^{(2)} \nonumber \\
\Rightarrow N_1 &= N_2. \label{lddw-n1-equals-n2}
\end{align}
Moreover, the domain wall condition in $T_2$ requires:
\begin{align}
\rho^{(2)}_\text{LD} + \rho^{(2)}_\text{HD} = 1 \nonumber \\
\Rightarrow \alpha_\text{eff}^{(2)} &= \beta_\text{eff}^{(2)} \nonumber \\
\Rightarrow \alpha \frac{N_2}{L} &= \beta \left(1 - \frac{N_1}{L} \right). \label{lddw-n1-and-n2}
\end{align}
Using Eq.~(\ref{lddw-n1-equals-n2}) in Eq.~(\ref{lddw-n1-and-n2}), we find:
\begin{equation}
\label{lddw-n1-n2}
\frac{N_1}{L} = \frac{N_2}{L} = \frac{\beta}{\alpha + \beta}.
\end{equation}

Substituting this back into the expressions for the effective rates [using Eqs.~(\ref{eff-rates-T1})–(\ref{f-and-g})], we obtain:
\begin{equation}
\label{lddw-eff-rates}
\alpha_\text{eff}^{(i)} = \beta_\text{eff}^{(i)} = \frac{\alpha \beta}{\alpha + \beta}, \quad \text{for } i=1,2.
\end{equation}

This leads to a contradiction. For $T_1$ to be in the LD phase, the necessary condition is $\alpha_\text{eff}^{(1)} < \beta_\text{eff}^{(1)}$, whereas for $T_2$ to be in the DW phase, we require $\alpha_\text{eff}^{(2)} = \beta_\text{eff}^{(2)}$. However, Eq.~(\ref{lddw-eff-rates}) shows that $\alpha_\text{eff}^{(1)} = \beta_\text{eff}^{(1)}$, which implies that $T_1$ must also be in the DW phase. Therefore, the LD-DW phase is not allowed within our model.

By the particle-hole symmetry, the DW-LD phase is also ruled out. A similar argument applies to the HD-DW and DW-HD phases, excluding them as well.
\end{itemize}

%%%%%%%%%%%%%%%%%%%%%%%%%

 \section{Phase boundaries and phase transitions}
 \label{pb}

 The phase diagrams of our model (Figs.~\ref{phase-diagrams-of-the-model} and \ref{ph-pd}) differ substantially from that of the open TASEP, or from other models of TASEPs with finite resources. 
 We identify the boundaries delineating different phases according to MFT. This can be achieved by equating the currents in corresponding phases.

 As the system transitions from the LD-LD phase to the MC-MC phase, the bulk density gradually increases, reaching \(\rho = 1/2\) at the LD-LD/MC-MC phase boundary. Substituting \(\rho_{\text{LD-LD}} = 1/2\) into Eq.~(\ref{rho-for-ld-ld-phase}) yields the condition for this boundary:
 \begin{equation}
  \alpha = \frac{1}{2\mu - 1}. \label{ldld-and-mcmc-boundary}
 \end{equation}
Similarly, the HD-HD/MC-MC boundary is obtained by setting \(\rho_{\text{HD-HD}} = 1/2\) in Eq.~(\ref{rho-for-hd-hd-phase}):
\begin{equation}
  \beta = \frac{1}{3 - 2\mu}. \label{hdhd-and-mcmc-boundary}
 \end{equation}
Boundaries~(\ref{ldld-and-mcmc-boundary}) and (\ref{hdhd-and-mcmc-boundary}) appear as straight lines parallel to the \(\beta\) and \(\alpha\) axes, respectively, see the phase diagrams in Figs.~\ref{phase-diagrams-of-the-model}(b)-\ref{phase-diagrams-of-the-model}(d) and Figs.~\ref{ph-pd}(a)-\ref{ph-pd}(b). The requirement \(\alpha, \beta > 0\) in Eqs.~(\ref{ldld-and-mcmc-boundary}) and (\ref{hdhd-and-mcmc-boundary}) imposes the following range on \(\mu\) for which these boundaries can exist:
\begin{equation}
  \label{mcmc-range}
  \frac{1}{2} < \mu < \frac{3}{2}.
\end{equation}

 At the boundary between LD-LD and DW-DW phases, domain walls are formed at the exit end of $T_{1}$ and $T_{2}$, wherein at the HD-HD and DW-DW phase boundary, domain walls are located at the entry end. One substitutes $x_{w}^{(1)}=x_{w}^{(2)}=1$ and $x_{w}^{(1)}=x_{w}^{(2)}=0$ in Eq.~(\ref{xw1-xw2-eq}) to obtain the following as boundaries between LD-LD and DW-DW phases, and HD-HD and DW-DW phases respectively:
\begin{eqnarray}
  &&\mu(\alpha+\beta)=\beta(1+\alpha), \label{ldld-and-dwdw-bundary}\\
  &&(\mu-1)(\alpha+\beta)=\beta(1-\alpha). \label{hdhd-and-dwdw-bundary}
 \end{eqnarray}
 Depending on the value of $\mu$, these boundaries can be either straight lines or curved ones, as shown in Figs.~\ref{phase-diagrams-of-the-model} and \ref{ph-pd}.
 
 We now discuss the nature of the phase transitions across these phase boundaries. The $\alpha$-$\beta$ phase space of our model can exhibit the presence of either two or four phases in different regions of the $\alpha$-$\beta$ plane, depending sensitively on the specific value of $\mu,\;0 \le \mu \le 2$. For $0 \le \mu \le 1/2$, only the LD-LD and DW-DW phases are present; while for $3/2 \le \mu \le 2$, only the HD-HD and DW-DW phases are observed. In contrast, for intermediate values ($1/2 \le \mu \le 3/2$), all four phases -- LD-LD, HD-HD, MC-MC, and DW-DW -- are observed in different regions of the $\alpha$-$\beta$ plane, the regions meeting at a common point in the $\alpha$-$\beta$ plane. Due to particle-hole symmetry described by (\ref{tr-1}) - (\ref{tr-3}), the phase diagram for $\mu > 1$ can be obtained from that for $\mu < 1$. Furthermore, considering the bulk densities as the order parameters, the  transitions between different phases -- for example, from LD-LD or HD-HD to MC-MC or DW-DW -- are continuous transitions, since the bulk density does not show any jump across any of the phase boundaries. These are summarized in Table~\ref{tab:ddw-phase-summary}.

Interestingly, Figs.~\ref{phase-diagrams-of-the-model} and \ref{ph-pd} reveal that, for $1/2 < \mu < 3/2$, all four phase boundaries intersect at a common point. Since the transitions across the different phase boundaries are all continuous transitions, the transitions across  this common meeting point are also continuous in nature. We call this a \textit{multicritical point}; see Refs.~\cite{driv3,driv4,krug-prl,ef-lktasep-prl} in this context, and Ref.~\cite{chaikin} for discussions on multicritical points in equilibrium systems. The coordinates of this multicritical point in the $\alpha$-$\beta$ phase diagram can be obtained by simultaneously solving the equations~(\ref{ldld-and-mcmc-boundary}), (\ref{hdhd-and-mcmc-boundary}), (\ref{ldld-and-dwdw-bundary}), and (\ref{hdhd-and-dwdw-bundary}):
\begin{equation}
  (\alpha_{c},\beta_{c})=\bigg( \frac{1}{2\mu - 1}, \frac{1}{3 - 2\mu} \bigg). \label{mult_coordinate}
\end{equation}

\begin{figure*}[htb]
 \includegraphics[width=\textwidth]{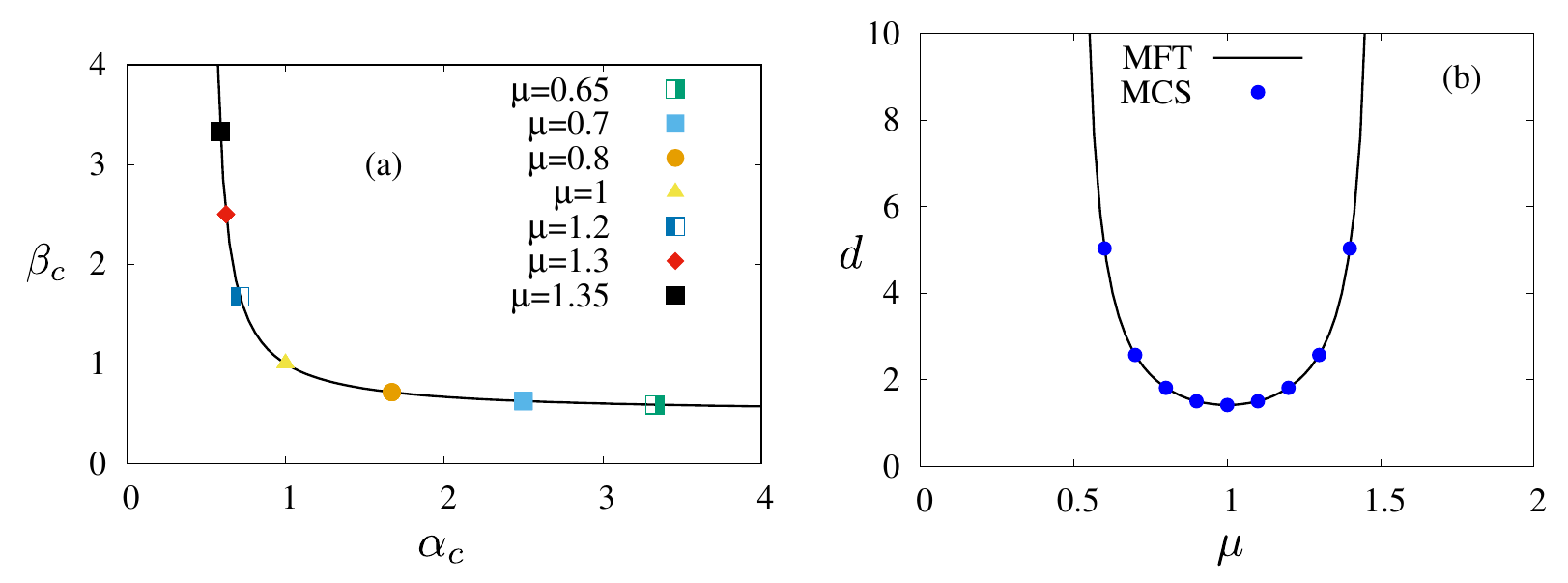}
 \\
\caption{(a) Hyperbolic trajectory of the multicritical point ($\alpha_{c},\beta_{c}$) for a set of representative values of $\mu$, see Eq.~(\ref{hyp_mult}). All four phase boundaries meet at a common point $(\alpha_{c},\beta_{c})=(1/(2\mu-1), 1/(3-2\mu))$ for $1/2 < \mu < 3/2$. (b) Variation of distance ($d$) between origin and multicritical point with $\mu$. Clearly, $d$ diverges as $\mu \rightarrow (1/2)^{+}$ and $\mu \rightarrow (3/2)^{-}$ and has a minima at half-filled limit $\mu=1$. Solid line represents the MF prediction which agrees well with the MCS results denoted by the discrete points.}
\label{two-figures}
\end{figure*}

As $\mu$ varies, the multicritical point~(\ref{mult_coordinate}) traces a hyperbolic trajectory [see Fig.~\ref{two-figures}(a)]:
\begin{equation}
  \alpha_c + \beta_c = 2 \alpha_c \beta_c. \label{hyp_mult}
\end{equation}

The distance \( d \) between the origin \((0,0)\) and the multicritical point \((\alpha_{c},\beta_{c})\) is given by:
\begin{equation}
\label{d}
  d = \sqrt{\frac{1}{(2\mu - 1)^2} + \frac{1}{(3 - 2\mu)^2}}.
\end{equation}
Notably, \(d\) diverges as \(\mu \rightarrow (1/2)^{+}\) and \(\mu \rightarrow (3/2)^{-}\), indicating that the multicritical point exists only within the range \(1/2 < \mu < 3/2\) [see Fig.~\ref{two-figures}(b)]. The symmetric nature of the plot of \(d\) versus \(\mu\) about \(\mu = 1\) reflects the particle-hole symmetry of the model.

%\begin{widetext}

%\begin{table}[h!]
%\centering
%\caption{Summary of the phases, domain walls, phase boundaries, and multicritical points as functions of $\mu$.}
%\label{tab:ddw-phase-summary}

%\begin{tabularx}{\textwidth}{|c|Y|Y|Y|Y|}
    %\hline
    %$\mu$ & Phases & Domain walls & Phase boundaries & Multicritical points \\
    %\hline
    %$0 \le \mu \le 1/2$ & LD-LD, DW-DW & A pair of DDWs & One second order & None \\
    %\hline
    %$1/2 \le \mu \le 3/2$ & LD-LD, HD-HD, MC-MC, DW-DW & A pair of DDWs & Four second order & One \\
    %\hline
    %$3/2 \le \mu \le 2$ & HD-HD, DW-DW & A pair of DDWs & One second order & None \\
    %\hline
%\end{tabularx}

%\end{table}

%\end{widetext}

 \begin{table*}
\centering
\renewcommand{\arraystretch}{1.5}
\caption{Summary of the phases, domain walls, phase boundaries, and multicritical points as functions of $\mu$.}
\label{tab:ddw-phase-summary}

\begin{tabular}{|c|c|c|c|c|}
\hline
$\mu$ & Phases & Domain walls & Phase boundaries & Multicritical points \\
\hline
$0 \le \mu \le 1/2$ & LD-LD, DW-DW & A pair of DDWs & One second order & None \\
\hline
$1/2 \le \mu \le 3/2$ & LD-LD, HD-HD, MC-MC, DW-DW & A pair of DDWs & Four second order & One \\
\hline
$3/2 \le \mu \le 2$ & HD-HD, DW-DW & A pair of DDWs & One second order & None \\
\hline
\end{tabular}

\end{table*}

 %\section{Delocalization of the domain walls}\label{ddw-details}

%{\bf why should plot (\ref{n1byn2-vs-mu}) be here? maybe it will go to the LD-LD phase section or in the appendix?}

%\begin{figure}[!h]
% \centering
% \includegraphics[width=0.5\textwidth]{n1byn2_vs_mu_a0.1_b2_crop.pdf}
% \caption{The plot shows the variation of the reservoir population ratio ($N_{1}/N_{2}$) as a function of $\mu$ in the LD-LD phase, see Eq.~(\ref{n1-n2-for-ld-ld-phase}). The control parameters are $\alpha=0.1$ and $\beta=2$. The results obtained from MFT and MCS exhibit good agreement.}
% \label{n1byn2-vs-mu}
% \end{figure}

%{%\bf can we generate a similar plot for the HD-HD phase?}

\section{CONCLUSION}
\label{conclusion}

In this work, we have investigated the stationary densities and domain walls in a model composed of two TASEP lanes connected antiparallelly to two particle reservoirs. This was originally proposed in Ref.~\cite{astik-erwin} as a model for TASEPs with distributed resources. Considering the same model but in a {\em different} reduced parameter space, where  the entry rates in both the TASEP lanes are equal and parametrized by $\alpha$ and the exit rates are in both lanes are also equal and parametrized by $\beta$. In addition, two functions $f$ and $g$ are defined in Eq.~(\ref{f-and-g}), which control the effective entry and exit rates, respectively. 
The exact quantitative results depend upon the forms of the functions $f$ and $g$ selected here, but even if $f$ and $g$ have different functional forms, the results will still be qualitatively  similar as long as they are increasing and decreasing monotonically with their arguments and the resources are finite. To keep things simple, the functions $f$ and $g$ are connected in a straightforward way and are presumed to have simple forms. This guarantees that $f$ increases monotonically with the reservoir population while $g$ decreases monotonically. We study the models by employing mean-field theories, supplemented by extensive Monte Carlo simulations. We calculate the stationary densities and the phase diagrams in terms of $\alpha,\,\beta$ and filling factor $\mu$.  

In spite of the simplifying reduction of the model, it shows hitherto unknown and complex phase behavior. For instance, the two TASEP lanes have the same phases --- LD-LD, HD-HD, MC-MC and DW-DW; distinct phases in the two TASEP lanes are {\em not} possible, unlike in the work of Ref.~\cite{astik-erwin}. This forms a major conclusion from this work. Furthermore, the phase diagrams are very different from those reported in Ref.~\cite{astik-erwin}. Remarkably, in the DW-DW phase this model admits domain walls in {\em each} of the TASEP lane only in the form of a pair of delocalized domain walls (DDWs), and not a single localized domain wall (LDW). Rather interestingly, the phase space region occupied by the DDW solutions covers an extended region. This is in stark contrast to other known models of TASEP, whether with open boundaries, or closed geometries or finite resources such as the one studied in Ref.~\cite{astik-erwin}. Indeed, all such other models of TASEP, the phase space region for a single or more than one DDWs is inevitably a line in the plane of the control parameters. Interestingly, the large number fluctuations in the TASEP lanes in the DW-DW phase is {\em not} accompanied by a corresponding large fluctuations in the reservoir particles numbers; the relative fluctuations in the latter actually should vanish in the thermodynamic limit. This is not just a theoretical novelty, it has a strong phenomenological significance as well. For instance, in possible practical realizations of this model, the presence of a pair of DDWs in an extended region of the phase space means that for an extended range of the control parameters (here, $\alpha,\,\beta$), the stationary densities are actually strongly fluctuating even in the thermodynamic limit (see Figs.~\ref{coarse-den-vs-time} and \ref{variance-of-den-vs-L}), since the DDWs necessarily involve mass fluctuations that scale with the system size (and hence survive in the thermodynamic limit).  Considering this as a simplified description for ribosome translation along messenger RNA strands in biological cells with a fixed number of ribosomes available, we expect for some ranges of the entry and exit rate parameters, there should be large fluctuations of the ribosome numbers along the messenger RNA strands. While any direct experimental results on such large fluctuations are not yet known, we expect our studies will provide new impetus to study along these lines. 
We have further studied the phase transitions, all of which are continuous in nature, with the bulk density appearing as the order parameter.  We have also calculated all of
the phase boundaries analytically within mean-field theory, which are in good agreement with the corresponding MCS results. Moreover, we have also identified the different threshold values of $\mu$ at
which different phase transitions occur in this model (see Table~\ref{tab:ddw-phase-summary}). Furthermore, we have calculated
the occupations $N_1$ and $N_2$ of the two reservoirs, which are always equal, independent of the phases on the TASEP lanes. Thus the model in this particular reduced parameter space chosen here does not allow any population imbalances in different parts of a bulk system.

We now briefly compare and contrast our results with those in Ref.~\cite{astik-erwin}. While the two models are structurally same and both admit the particle-hole symmetry, the reduced parameter space chosen here is distinct from the one considered in Ref.~\cite{astik-erwin}. As a result, the phase diagrams in the present study are quite different from those reported in Ref.~\cite{astik-erwin}. The two TASEP lanes have statistically identical stationary densities here, unlike in Ref.~\cite{astik-erwin}, where distinct stationary densities on the two TASEP lanes are also possible. Related to this, the two reservoirs always have the same steady state occupations in the present study, ruling out any macroscopic population imbalances, which differs from Ref.~\cite{astik-erwin}. Furthermore, all the phase boundaries in this study correspond to continuous transitions, whereas discontinuous transitions were also found in Ref.~\cite{astik-erwin}. Lastly and as mentioned above, the present model admits a pair of DDWs over extended regions of the parameter space, whereas for model in Ref.~\cite{astik-erwin}, a pair of DDWs are possible only for the control parameters falling on lines in the phase space. On the other hand, there are no localized domain walls in the present model, which are found in Ref.~\cite{astik-erwin} for the control parameters in extended regions of the phase space.

 We note that in spite of the adhoc nature of the MFT, it agrees very well with the MCS results, a fact also holds for Ref.~\cite{astik-erwin}. This is not generally expected as the overall particle number conservation should make the density fluctuations correlated, which is neglected in MFT.  This unexpected agreement between the MFT and MCS results may be heuristically understood by considering that  the presence of the attached particle reservoirs ensure that there are {\em no} conservation laws associated with the individual TASEP lanes, making them similar to open TASEPs. The latter have no conservation laws, and consequently, the density fluctuations are uncorrelated, making the MFT a good and accurate described. The presence of the particle reservoirs similarly make the density fluctuations in the TASEP lane uncorrelated, which makes the MFT accurate. See also Ref.~\cite{def5} similar discussions in a related context.  

This work can be extended in various ways. For instance, one can increase the number of TASEP lanes and number of reservoirs. Our above framework including our mean-field analysis can be easily extended to any
number of reservoirs and TASEP lanes with equal entry and exit rates (as here). Nonetheless, we expect the broad qualitative
features observed in our study to persist. For instance,  it is reasonable to expect that all the channels are to be in their LD (HD) phases for very low (high) availability of resources $\mu$. As $\mu$ rises DW and MC phases should appear. It is expected that these DWs should all be delocalized for reasons similar to the above. One can also consider using reservoirs having functions defining the entry and exit rates that are different from the ones used here.  These qualitative predictions  can be confirmed by stochastic simulation studies. Furthermore, making the parameters $\alpha,\,\beta$, which parametrize the effective entry and exit rates, fluctuate stochastically with the specified distributions would be still another intriguing extension. Our study here should serve as a catalyst for further investigation in these directions.

{\em Acknowledgment:-} A.B. thanks the AvH Stiftung
(Germany) for partial financial support under the Research
Group Linkage Programme scheme (2024).

 \appendix
 
 \section{Particle-hole symmetry and phase diagrams for $1 < \mu \le 2$}
 \label{ph-and-pd}

A particle-hole symmetry is embedded in the dynamics of our model due to the specific choice of the rate functions \( f \) and \( g \) [see Eq.~\eqref{f-and-g}]. This symmetry operates as follows: the absence of a particle at a site corresponds to the presence of a hole, and the hopping of a particle in one direction is equivalent to the hopping of a hole in the opposite direction. Consequently, the entry of a particle into a TASEP lane corresponds to the exit of a hole from that lane at the same rate, and vice versa.

For clarity, we rewrite the equations of motion for the entry site (\( j=1 \)), bulk sites (\( 1 < j < L \)), and exit site (\( j = L \)) of the TASEP lanes \( T_i \), with \( i = 1, 2 \):
\begin{align}
\partial_t \rho_1^{(i)} &= \alpha_\text{eff}^{(i)} (1 - \rho_1^{(i)}) - \rho_1^{(i)} (1 - \rho_2^{(i)}), \label{entry-eqn} \\
\partial_t \rho_j^{(i)} &= \rho_{j-1}^{(i)} (1 - \rho_j^{(i)}) - \rho_j^{(i)} (1 - \rho_{j+1}^{(i)}), \label{bulk-eqn} \\
\partial_t \rho_L^{(i)} &= \rho_{L-1}^{(i)} (1 - \rho_L^{(i)}) - \beta_\text{eff}^{(i)} \rho_L^{(i)}. \label{exit-eqn}
\end{align}
We now consider the following transformations:
\begin{align}
\alpha_i &\longleftrightarrow \beta_i, \label{tr1} \\
\rho_j^{(i)} &\longleftrightarrow 1 - \rho_{L - j + 1}^{(i)}, \label{tr2} \\
\mu &\longleftrightarrow 2 - \mu. \label{tr3}
\end{align}
Under transformation~\eqref{tr2}, the terms in the entry and exit equations [Eqs.~\eqref{entry-eqn} and \eqref{exit-eqn}, respectively] involving nearest-neighbor hopping are interchanged:
\[
\rho_1^{(i)} (1 - \rho_2^{(i)}) \;\longleftrightarrow\; \rho_{L-1}^{(i)} (1 - \rho_L^{(i)}),
\]
and the site densities at the boundaries transform as
\[
\rho_1^{(i)} \;\longleftrightarrow\; 1 - \rho_L^{(i)}.
\]
Next, we analyze how the effective boundary rates transform. From particle number conservation and using the filling factor \( \mu = N_0 / 2L \), we have:
\begin{equation}
\label{pnc-again}
2\mu = \frac{N_1}{L} + \frac{N_2}{L} + \frac{1}{L} \sum_{i=1}^2 \sum_{j=1}^L \rho_j^{(i)}.
\end{equation}
Applying transformation~\eqref{tr2}, the density sum transforms as
\[
\sum_{j=1}^L \rho_j^{(i)} \;\longleftrightarrow\; L - \sum_{j=1}^L \rho_j^{(i)},
\]
and together with \eqref{tr3}, the reservoir terms satisfy:
\[
\frac{N_1}{L} + \frac{N_2}{L} \;\longleftrightarrow\; \left(1 - \frac{N_2}{L}\right) + \left(1 - \frac{N_1}{L}\right),
\]
which implies the component-wise transformations:
\[
\frac{N_1}{L} \;\longleftrightarrow\; 1 - \frac{N_2}{L}, \qquad \frac{N_2}{L} \;\longleftrightarrow\; 1 - \frac{N_1}{L}.
\]
Substituting into the expressions for the effective rates:
\begin{align*}
\alpha_\text{eff}^{(1)} = \alpha_1 \frac{N_1}{L} &\;\longleftrightarrow\; \beta_1\left(1 - \frac{N_2}{L}\right) = \beta_\text{eff}^{(1)}, \\
\alpha_\text{eff}^{(2)} = \alpha_2 \frac{N_2}{L} &\;\longleftrightarrow\; \beta_2\left(1 - \frac{N_1}{L}\right) = \beta_\text{eff}^{(2)}.
\end{align*}
Thus, under the transformations~\eqref{tr1}--\eqref{tr3}, the entry-site equation~\eqref{entry-eqn} maps to the exit-site equation~\eqref{exit-eqn}, demonstrating a symmetry between particle entry and hole exit. For bulk sites, Eq.~\eqref{bulk-eqn} transforms into:
\[
\partial_t \rho_{L - j + 1}^{(i)} = \rho_{L - j}^{(i)} \left(1 - \rho_{L - j + 1}^{(i)}\right) - \rho_{L - j + 1}^{(i)} \left(1 - \rho_{L - j + 2}^{(i)}\right),
\]
confirming that the bulk dynamics are invariant under the particle-hole transformation.

\begin{figure*}[htb]
\centering
\includegraphics[width=0.32\textwidth]{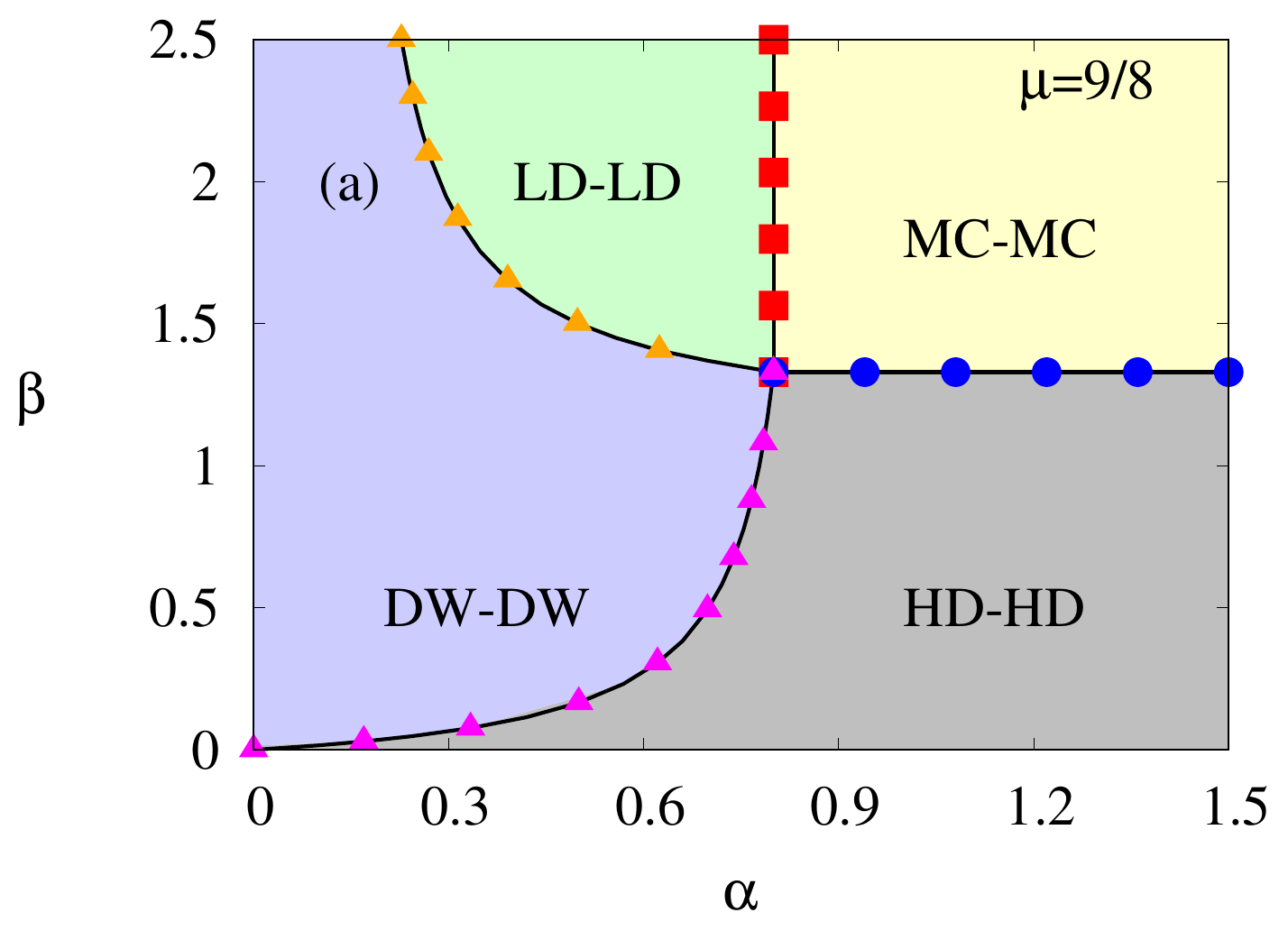}
\hfill
\includegraphics[width=0.32\textwidth]{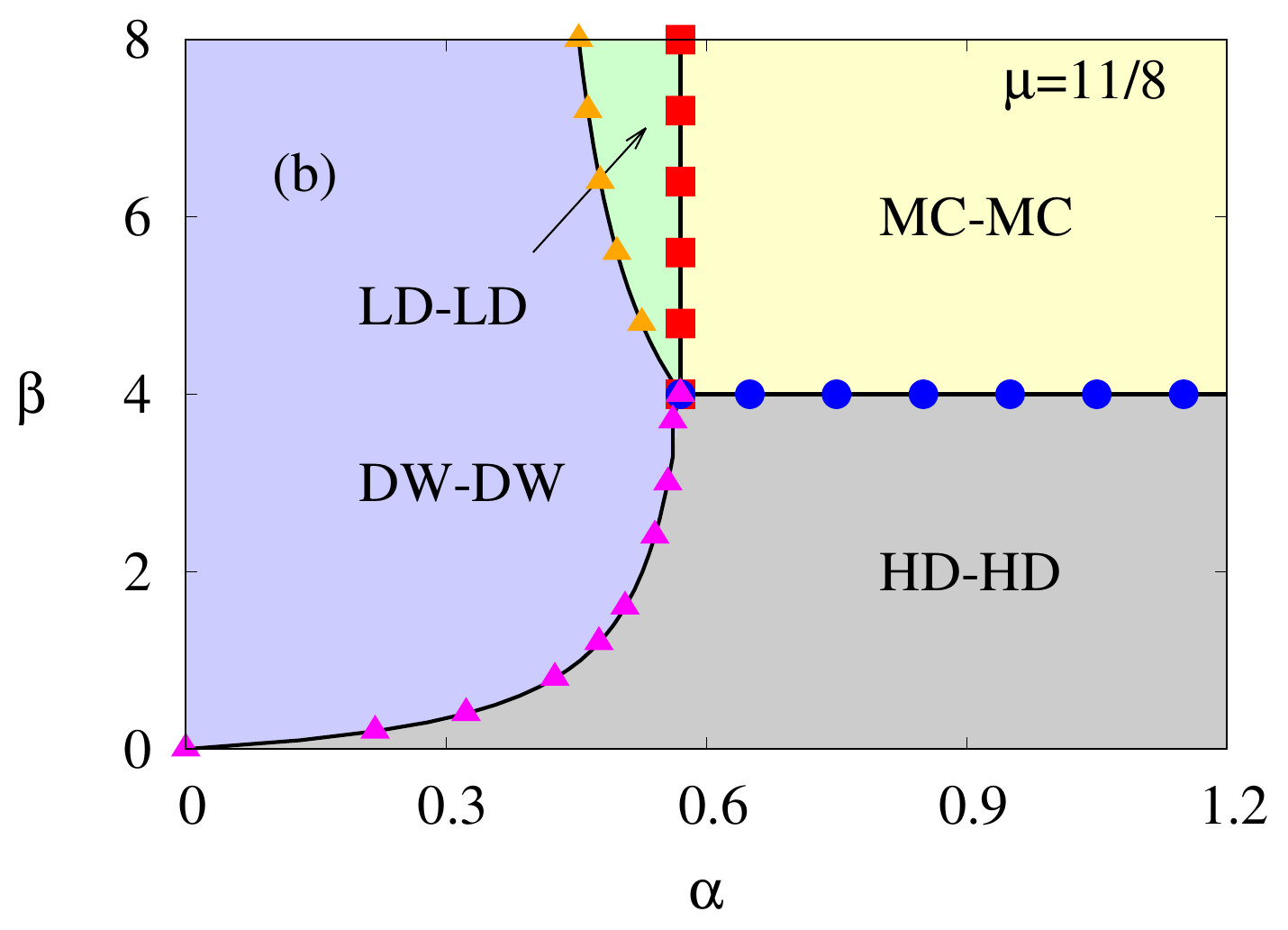}
\hfill
\includegraphics[width=0.32\textwidth]{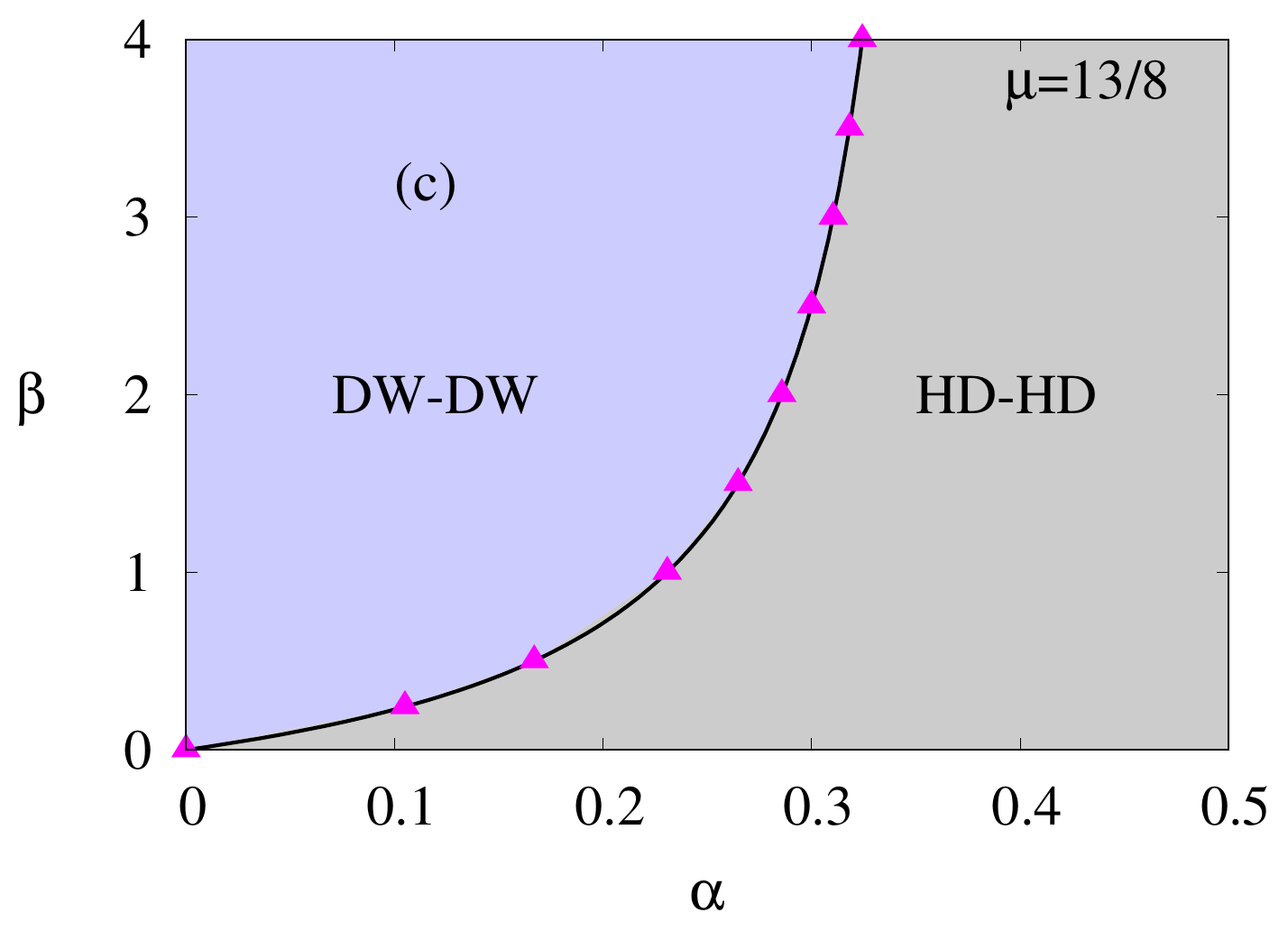}
\caption{Phase diagrams of the model in the $\alpha$--$\beta$ plane for $\mu > 1$ are shown. The phase diagrams for $\mu = 9/8$, $11/8$, and $13/8$ exhibit particle-hole symmetry with those corresponding to $\mu = 7/8$, $5/8$, and $3/8$, respectively (cf. Fig.~\ref{phase-diagrams-of-the-model}). System size is $L = 1000$ and $2 \times 10^9$ Monte Carlo steps are used. MCS results (discrete colored points) agree well with MFT predictions (black solid lines).}
\label{ph-pd}
\end{figure*}

We now demonstrate that the particle-hole symmetry is reflected in the phase diagrams of our model. In the main text, we presented phase diagrams for \( \mu \leq 1 \); see Fig.~\ref{phase-diagrams-of-the-model}. Here, we extend the analysis to the complementary regime \( 1 < \mu \leq 2 \), which—by symmetry—can be inferred from the previously obtained phase diagrams corresponding to \( \mu \leq 1 \) via the transformations~(\ref{tr1})--(\ref{tr3}). Figures~\ref{ph-pd}(a) (\( \mu = 9/8 \)), \ref{ph-pd}(b) (\( \mu = 11/8 \)), and \ref{ph-pd}(c) (\( \mu = 13/8 \)) are related, respectively, to Figs.~\ref{phase-diagrams-of-the-model}(c) (\( \mu = 7/8 \)), \ref{phase-diagrams-of-the-model}(b) (\( \mu = 5/8 \)), and \ref{phase-diagrams-of-the-model}(a) (\( \mu = 3/8 \)) by the transformations~(\ref{tr1})--(\ref{tr3}).

 \section{Steady state density profiles in the LD-LD, HD-HD, and MC-MC phases}
 \label{ld-hd-mc-den}

\begin{figure}[!h]
 \centering
 \includegraphics[width=0.5\textwidth]{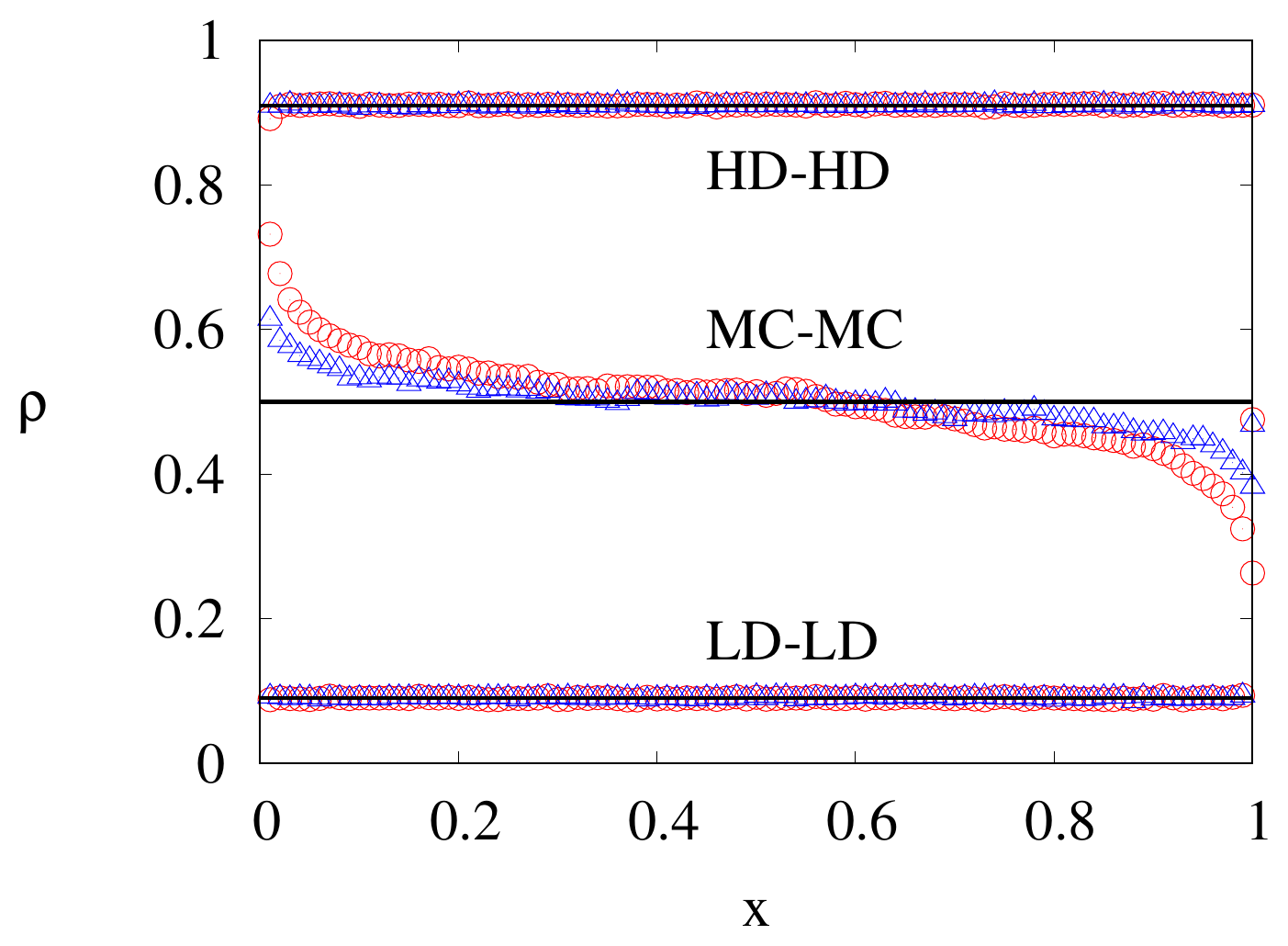}
 \caption{Steady state density profiles in the LD-LD, MC-MC, and HD-HD phases of the model. Filling factor is taken as $\mu=1$. Entry and exit rate parameters are: $\alpha=0.1$, $\beta=1.5$ for LD-LD phase; $\alpha=\beta=2$ for MC-MC phase; and $\alpha=1.5$, $\beta=0.1$ for HD-HD phase. System size is $L=1000$ and $2 \times 10^{9}$ Monte Carlo steps are taken. For both lanes, MCS densities (red circles for $T_{1}$ and blue triangles for $T_{2}$) support corresponding MFT predictions (black solid lines) very well.}
 \label{ld-hd-mc-den-profiles}
 \end{figure}

 Stationary state density profiles in the LD-LD, HD-HD, and MC-MC phases are presented in Fig.~\ref{ld-hd-mc-den-profiles}.  Both MFT and MCS studies results are shown, which mutually agree well.

\section{Detailed calculations of the HD-HD phase}
 \label{hdhd-detailed}

In this section, we explicitly derive the effective entry and exit rates in the HD-HD phase, determine the extent of the HD-HD region in the $\alpha$-$\beta$ plane, and obtain the stationary density characterizing the HD-HD phase.

%%%%%%%%%%%%%%%%%%%%%
The HD-HD phase also corresponds to the solution $\rho^{(1)} = 1 - \beta_\text{eff}^{(1)} = \rho^{(2)} = 1 - \beta_\text{eff}^{(2)} \equiv \rho_\text{HD-HD} > 1/2$. This yields $N_{1}=N_{2}$ similar to the LD-LD phase:
\begin{align}
&1 - \beta_\text{eff}^{(1)} = 1 - \beta_\text{eff}^{(2)} \nonumber \\
\implies\quad &1 - \beta \left(1 - \frac{N_2}{L} \right) = 1 - \beta \left(1 - \frac{N_1}{L} \right) \nonumber \\
\implies\quad &N_1 = N_2. \label{hdhd-n1-n2}
\end{align}
The PNC condition then determines $N_1$ and $N_2$ in terms of the control parameters $\beta$ and $\mu$:
\begin{equation}
\label{n1-n2-for-hd-hd-phase-apndx}
\frac{N_1}{L} = \frac{\mu - 1 + \beta}{1 + \beta} = \frac{N_2}{L}.
\end{equation}
Again, $N_1$ and $N_2$ scale linearly with the system size $L$, and depend linearly on the filling factor $\mu$.
%%%%%%%%%%%%%%%%%%%%%%%%%%555

Since $0 < N_1,N_2 \le L$, Eq.~(\ref{n1-n2-for-hd-hd-phase-apndx}) yields
\begin{equation}
\label{hdhd-ineq-1-apndx}
 \mu \le 2 \quad \text{and} \quad \frac{2-\mu}{1+\beta} < 1.
\end{equation}
Substituting Eq.~(\ref{n1-n2-for-hd-hd-phase-apndx}) in Eqs.~(\ref{eff-rates-T1})-(\ref{f-and-g}) one gets the effective entry and exit rates in the HD-HD phase:
\begin{eqnarray}
  &&\alpha_\text{eff}^{(1)}=\alpha_\text{eff}^{(2)}=\frac{\alpha(\mu-1+\beta)}{1+\beta}, \label{alphaeff-hdhd-apndx}\\
  &&\beta_\text{eff}^{(1)}=\beta_\text{eff}^{(2)}=\frac{(2-\mu)\beta}{1+\beta}. \label{betaeff-hdhd-apndx}
 \end{eqnarray}
Analogous to an open TASEP, the effective rates in the HD-HD phase obtained in Eqs.~(\ref{alphaeff-hdhd-apndx}) and (\ref{betaeff-hdhd-apndx}) must satisfy $\beta_\text{eff}^{(i)} < \alpha_\text{eff}^{(i)}$
and $\beta_\text{eff}^{(i)} < 1/2$ for $i=1,2$ resulting in the following two inequalities:
\begin{eqnarray}
  &&\frac{2-\mu}{1+\beta} < \frac{\alpha}{\alpha+\beta}, \label{hdhd-ineq-2-apndx}\\
  &&\frac{(2-\mu)\beta}{1+\beta} < \frac{1}{2}. \label{hdhd-ineq-3-apndx}
 \end{eqnarray}
 Inequalities~(\ref{hdhd-ineq-1-apndx}), (\ref{hdhd-ineq-2-apndx}), and (\ref{hdhd-ineq-3-apndx}) together define the HD-HD region in the $\alpha$-$\beta$ plane for a fixed $\mu$. These conditions are identical to those obtained via particle-hole symmetry, Eqs.~(\ref{hdhd-ineq-1}), (\ref{hdhd-ineq-2}), and (\ref{hdhd-ineq-3}).

 The steady-state densities in both lanes in the HD-HD phase are obtained as
\begin{equation}
\label{rho-for-hd-hd-phase-apndx}
\rho_\text{HD-HD} = 1-\beta_\text{eff}^{(1(,2))} = \frac{1 - \beta + \mu \beta}{1 + \beta}.
\end{equation}


\begin{thebibliography}{99}

\bibitem{driv1} S. Katz, J. L. Lebowitz, and H. Spohn, Phase transitions in stationary nonequilibrium states of model lattice systems, Phys. Rev. B 28, 1655 (1983). 

\bibitem{driv2} S. Katz, J. L. Lebowitz, and H. Spohn, Nonequilibrium steady states of stochastic lattice gas models of fast ionic conductors, J. Stat. Phys. 34 497 (1984).

\bibitem{driv3} B. Schmittmann and R. K.-P. Zia, in {\em Phase Transitions and Critical Phenomena}, edited by J. L. Lebowitz and C. Domb, Vol. 17 (Academic Press, London, 1995).

\bibitem{driv4} {\em Nonequilibrium Statistical Mechanics in One Dimension}, edited by V. Privman (Cambridge University Press, Cambridge, 1997).

\bibitem{krug-prl} J. Krug, Phys. Rev. Lett. {\bf 67}, 1882 (1991).

\bibitem{derrida1} B. Derrida, S. A. Janowsky, J. L. Lebowitz, and E. R. Speer, Exact solution of the totally asymmetric simple exclusion process:
Shock profiles, J. Stat. Phys. {\bf 73}, 813 (1993).

\bibitem{derrida2} B. Derrida and M. R Evans, Exact correlation functions in an asymmetric exclusion model with open boundaries, J. Phys. I 3, 311 (1993).

\bibitem{derrida3} B. Derrida, M. R. Evans, V. Hakim, and V. Pasquier, Exact solution of a 1D asymmetric exclusion model using a matrix formulation, J. Phys. A: Math. Gen. {\bf 26}, 1493 (1993).

%\bibitem{derrida4}

\bibitem{macdonald} C. T. MacDonald, J. H. Gibbs, and A. C. Pipkin, ``Kinetics of biopolymerization on nucleic acid templates'', Biopolymers {\bf 6}, 1 (1968).

\bibitem{ef-lktasep-prl}  A. Parmeggiani, T. Franosch, and E. Frey, Phase coexistence in
driven one-dimensional transport, Phys. Rev. Lett. {\bf 90}, 086601
(2003).

\bibitem{ef-lktasep-pre} A. Parmeggiani, T. Franosch, and E. Frey, Totally asymmetric
simple exclusion process with Langmuir kinetics, Phys. Rev. E {\bf 70}, 046101 (2004).

\bibitem{traffic-exp} C. Leduc, K. Padberg-Gehle, V. Varga, D. Helbing, S. Diez, and
J. Howard, Molecular crowding creates traffic jams of kinesin motors on microtubules, Proc. Natl. Acad. Sci. USA {\bf 109}, 6100
(2012).

\bibitem{melbinger1} A. Melbinger, L. Reese, and E. Frey, Microtubule length regulation by molecular motors, Phys. Rev. Lett. {\bf 108}, 258104
(2012).

\bibitem{melbinger2} L. Reese, A. Melbinger, and E. Frey, Crowding of molecular motors determines microtubule depolymerization, Biophys. J. {\bf 101}, 2190 (2011).

\bibitem{reichenbach1} T. Reichenbach, T. Franosch, and E. Frey, Exclusion processes
with internal states, Phys. Rev. Lett. {\bf 97}, 050603 (2006).

\bibitem{reichenbach2} T. Reichenbach, E. Frey, and T. Franosch, Traffic jams induced by rare switching events in two-lane transport, New J. Phys. {\bf 9}, 159 (2007).

\bibitem{reichenbach3} T. Reichenbach, T. Franosch, and E. Frey, Domain wall delocalization, dynamics and fluctuations in an exclusion process with two internal states, Eur. Phys. J. E {\bf 27}, 47 (2008).

\bibitem{net1}  B. Embley, A. Parmeggiani, and N. Kern, Understanding totally asymmetric simple-exclusion-process transport on networks:
Generic analysis via effective rates and explicit vertices, Phys. Rev. E {\bf 80}, 041128 (2009).

\bibitem{net2} R. Chatterjee, A. K. Chandra, and A. Basu, Phase transition and phase coexistence in coupled rings with driven exclusion processes, Phys. Rev. E {\bf 87}, 032157 (2013).

\bibitem{net3} R. Chatterjee, A. K. Chandra, and A. Basu, Asymmetric exclusion processes on a closed network with bottlenecks, J. Stat.
Mech.: Theory Exp. (2015) P01012.

\bibitem{net-exp} J. Howard and R. L. Clark, Mechanics of motor proteins and the cytoskeleton, Appl. Mech. Rev. {\bf 55}, B39 (2002).

\bibitem{def1} S. A. Janowsky and J. L. Lebowitz, Finite-size effects and shock fluctuations in the asymmetric simple-exclusion process, Phys. Rev. A {\bf 45}, 618 (1992).

\bibitem{def2} H. Hinsch and E. Frey, Bulk-driven nonequilibrium phase transitions in a mesoscopic ring, Phys. Rev. Lett. {\bf 97}, 095701 (2006).

\bibitem{sarkar-basu} N. Sarkar and A Basu, Nonequilibrium steady states in asymmetric exclusion processes on a ring with bottlenecks, Phys. Rev. E {\bf 90}, 022109 (2014).

\bibitem{banerjee-sarkar-basu} T. Banerjee, N. Sarkar, and A. Basu, Generic nonequilibrium
steady states in an exclusion process on an inhomogeneous ring, J. Stat. Mech.: Theory Exp. (2015) P01024.

\bibitem{def5} T. Banerjee, A. K. Chandra, and A. Basu, Phase coexistence and particle nonconservation in a closed asymmetric exclusion
process with inhomogeneities, Phys. Rev. E {\bf 92}, 022121 (2015).

\bibitem{def6} B. Daga, S. Mondal, A. K. Chandra, T. Banerjee, and A. Basu, Nonequilibrium steady states in a closed inhomogeneous asymmetric exclusion process with generic particle nonconservation, Phys. Rev. E {\bf 95}, 012113 (2017).

\bibitem{def7} T. Banerjee and A. Basu, Smooth or shock: Universality in closed inhomogeneous driven single file motions, Phys. Rev. Res. {\bf 2}, 013025 (2020).

\bibitem{def8} P. Roy, A. K. Chandra, and A. Basu, Pinned or moving: States of a single shock in a ring, Phys. Rev. E {\bf 102}, 012105 (2020).

\bibitem{def9} A. Goswami, R. Chatterjee, and S. Mukherjee, Defect versus defect: Stationary states of single file marching in periodic landscapes with road blocks, Phys. Rev. E {\bf 110}, 064149
(2024).

%\bibitem{sm-rohn-atri} A. Goswami, R. Chatterjee, and S. Mukherjee, Defect
%versus defect: stationary states of single file marching in
%periodic landscapes with road blocks, Phys. Rev. E {\bf 110}, 064149 (2024).

\bibitem{def10} S. Pal and A. Basu, Are shocks in conserved driven single-file motions with bottlenecks universal? arXiv:  2505.09497.

\bibitem{half1} I. R. Graf and E. Frey, Generic transport mechanisms for molecular traffic in cellular protrusions, Phys. Rev. Lett. {\bf 118}, 128101
(2017).

\bibitem{half2} M. Bojer, I. R. Graf, and E. Frey, Self-organized system-size oscillation of a stochastic lattice-gas model, Phys. Rev. E {\bf 98},
012410 (2018).

\bibitem{half3} A. Goswami, U. Dey, and S. Mukherjee, Nonequilibrium steady states in coupled asymmetric and symmetric exclusion processes, Phys. Rev. E {\bf 108}, 054122 (2023).

\bibitem{klumpp1} R. Lipowsky, S. Klumpp, and T. M. Nieuwenhuizen, Random walks of cytoskeletal motors in open and closed compartments, Phys. Rev. Lett. {\bf 87}, 108101 (2001).

\bibitem{klumpp2} S. Klumpp and R. Lipowsky, Traffic of molecular motors through tube-like compartments, J. Stat. Phys. {\bf 113}, 233 (2003).

\bibitem{ciandrini} L. D. Fernandes and L. Ciandrini, Driven transport on a flexible polymer with particle recycling: A model inspired by transcription and translation, Phys. Rev. E {\bf 99}, 052409 (2019).

\bibitem{dauloudet} O. Dauloudet, I. Neri, J.-C. Walter, J. Dorignac, F. Geniet, and A. Parmeggiani, Modelling the effect of ribosome mobility on the rate of protein synthesis, Eur. Phys. J. E {\bf 44}, 1 (2021).

\bibitem{klumpp3} S. Klumpp and R. Lipowsky, Asymmetric simple exclusion processes with diffusive bottlenecks, Phys. Rev. E {\bf 70}, 066104
(2004).

\bibitem{Chowdhury-Santen-Schadschneider} D. Chowdhury, L. Santen, and A. Schadschneider, ``Statistical physics of vehicular traffic and
some related systems'', Phys. Rep., {\bf  329}, Nos. 4-6 (2000).

\bibitem{Schadschneider} A. Schadschneider, ``Traffic flow: a statistical physics point of view'', Physica A, {\bf 329}, Nos. 4-6 (2002).

\bibitem{reser1} D. A. Adams, B. Schmittmann, and R. K. P. Zia, Far-from-equilibrium transport with constrained resources, J. Stat. Mech.: Theory Exp. (2008) P06009.

\bibitem{ha-den} M. Ha and M. den Nijs, Macroscopic car condensation in a parking garage, Phys. Rev. E {\bf 66}, 036118 (2002).

\bibitem{traff1} C. A. Brackley, M. C. Romano, C. Grebogi, and M. Thiel, Limited resources in a driven diffusion process, Phys. Rev. Lett. {\bf 105}, 078102 (2010).

\bibitem{traff2} C. A. Brackley, M. C. Romano, and M. Thiel, Slow sites in an exclusion process with limited resources, Phys. Rev. E {\bf 82}, 051920 (2010).


\bibitem{reser2} L. Jonathan Cook, R. K. P. Zia, Feedback and fluctuations in a totally asymmetric simple exclusion process
with finite resources, J. Stat. Mech.: Theory Exp 2009,
P02012 (2009).

\bibitem{reser3} L. Jonathan Cook, R. K. P. Zia, and B. Schmittmann,
Competition between multiple totally asymmetric simple exclusion processes for a finite pool of resources,
Phys. Rev. E 80, 031142 (2009).



\bibitem{cook-dong-lafleur} L. J. Cook, J. J. Dong, and A. LaFleur, Interplay between finite resources and a local defect in an asymmetric simple exclusion process, Phys. Rev. E {\bf 88}, 042127 (2013).

\bibitem{astik-parna} A Haldar, P Roy, A Basu, Asymmetric exclusion processes with fixed resources: Reservoir crowding and steady states, Phys. Rev. E {\bf 104}, 034106 (2021).

%\bibitem{arvind-dynamic-defect} 

%\bibitem{arvind-2}
\bibitem{arvind2} N. Bhatia and A. K. Gupta, Role of site-wise dynamic defects in a resource-constrained exclusion process, Chaos, Solitons and Fractals {\bf 167}, 113109 (2023).


\bibitem{sourav-1} S. Pal, P. Roy and A. Basu, ``Availability, storage capacity, and diffusion: Stationary states of an asymmetric exclusion process connected to two reservoirs'', Phys. Rev. E {\bf 110}, 054104 (2024).

\bibitem{sourav-2} S. Pal, P. Roy and A. Basu, ``Distributed fixed resources exchanging particles: Phases of an asymmetric exclusion process connected to two reservoirs'', Phys. Rev. E {\bf 111}, 034109 (2025).

\bibitem{astik-erwin} A. Haldar, P. Roy, E. Frey and A. Basu, ``Availability versus carrying capacity: Phases of asymmetric exclusion processes competing for finite pools of resources'', Phys. Rev. E {\bf 111}, 014154 (2025).

\bibitem{arvind-bi} A. Gupta, B. Pal, and A. K. Gupta, Interplay of reservoirs in a bidirectional system, Phys. Rev. E 107, 034103 (2023).

\bibitem{seppa} T. Sepp\"al\"inen, Existence of hydrodynamics for the totally asymmetric exclusion simple K-exclusion process, Ann. Probab. {\bf 27}, 361 (1999).

%\bibitem{sourav-4} S. Pal and A. Basu, Universality of shocks in conserved driven single-file motions with
 %   bottlenecks, arXiv: 

\bibitem{blythe} R A Blythe and M R Evans, ``Nonequilibrium steady states of matrix-product form: a solver's guide'', J. Phys. A {\bf 40}, R333 (2007).


\bibitem{greulich-ciandrini} P. Greulich, L. Ciandrini, R. J. Allen, and M. C. Romano, ``Mixed population of competing totally asymmetric simple exclusion processes with a shared reservoir of particles'', Phys. Rev. E {\bf 85}, 011142 (2012).
 
%\bibitem{adams-schmittmann-zia} D. A. Adams, B. Schmittmann, and R. K. P. Zia, ``Far-from-equilibrium transport with constrained resources'', J. Stat. Mech.: Theory Exp 2008, P06009 (2008).

\bibitem{arvind1} B. Pal and A. K. Gupta, Reservoir crowding in a resource-constrained exclusion process with a dynamic defect, Phys. Rev. E {\bf 106}, 044130 (2022).


%\bibitem{schmittmann-zia}

\bibitem{chou-mallick-zia} T. Chou, K. Mallick, and R. K. P. Zia, Rep. Prog. Phys. {\bf 74}, 116601 (2011).

\bibitem{tasep-rev} T. Chou, K. Mallick, and R. K. P. Zia, ``Non-equilibrium statistical mechanics: from a paradigmatic model to biological transport'', Rep. Prog. Phys. {\bf 74}, 116601 (2011).

\bibitem{rajewsky-update}
N. Rajewsky, L. Santen, A. Schadschneider, and M. Schreckenberg,
The asymmetric exclusion process: Comparison of update procedures,
J. Stat. Phys. {\bf 92}, 151 (1998).

\bibitem{schreckenberg-traffic}
M. Schreckenberg, A. Schadschneider, K. Nagel, and N. Ito,
Discrete stochastic models for traffic flow,
Phys. Rev. E {\bf 51}, 2939 (1995).

\bibitem{sm-inhomogeneous-tasep} S. Mukherjee and A. Basu, Nonuniform asymmetric exclusion process: Stationary densities and domain walls, arXiv: 2410.22516 (2024).

\bibitem{chaikin} P. M. Chaikin and T. C. Lubensky, Principles of Condensed
Matter Physics (Cambridge University Press, Cambridge, UK,
2000).
\end{thebibliography}
\end{document}